\documentclass[12pt]{article}
\usepackage{amsmath,amssymb,enumerate}

   {\end{list}}%

\makeatletter
\renewcommand{\section}{\setcounter{equation}{0}\@startsection
  {section}%
  {1}%
  {0pt}%
  {-1\baselineskip}%
  {0.4\baselineskip}%
  {\bfseries\large}}%
\renewcommand{\subsection}{\@startsection
  {subsection}%
  {2}%
  {0pt}%
  {-0.75\baselineskip}%
  {0.2\baselineskip}%
  {\bfseries}}%
\renewcommand{\subsubsection}{\@startsection
  {subsubsection}%
  {3}%
  {0pt}%
  {-0.5\baselineskip}%
  {0.1\baselineskip}%
  {\sc}}%
\makeatother

\def\m{\mu}
\def\n{\nu}
\def\r{\rho}
\def\s{\sigma}

\def\id{{\rm{I}\!\rm{I}}}
 \def\tr{\text{Tr}}
 
\def\id3x{\int\!\! d^3\!\vec{x}}
\def\idx{\int\!\! d^4\!x}

\def\rig>{\right>}

\newcommand{\bea}{\begin{eqnarray}}
\newcommand{\eea}{\end{eqnarray}}
\newcommand{\beann}{\begin{eqnarray*}}
\newcommand{\eeann}{\end{eqnarray*}}
\newcommand{\ba}{\begin{array}}
\newcommand{\ea}{\end{array}}

\newcommand{\Tr}{\mathbf{Tr}}
\newcommand{\ST}{\star}

\def\g5{\gamma_{5}}

\def\idx3{\int\! d^{3}\!\vec{x}\,}
\def\idx{\int\! d^{4}\!x\,}

 \def\am {a_{\mu}}

 \def\g {\gamma}

\def\r {\rho}
 \def\s {\sigma}

 \def\Tr{\text{Tr}}
\def\tr{\text{tr}}

\begin{document}

\begin{titlepage}
\rightline{FTI/UCM 80-2006} \vglue 45pt

\begin{center}

{\Large \bf  Monopoles,  noncommutative gauge
theories\\ in the BPS limit and some simple gauge groups.}\\
\vskip 1.2 true cm {\rm C.P. Mart\'{\i}n}\footnote{E-mail:
carmelo@elbereth.fis.ucm.es}
 and C. Tamarit\footnote{E-mail: ctamarit@fis.ucm.es}
\vskip 0.3 true cm {\it Departamento de F\'{\i}sica Te\'orica I,
Facultad de Ciencias F\'{\i}sicas\\
Universidad Complutense de Madrid,
 28040 Madrid, Spain}\\
\vskip 0.75 true cm

\vskip 0.25 true cm

{\leftskip=50pt \rightskip=50pt \noindent For three conspicuous
gauge groups, namely, $SU(2)$, $SU(3)$ and $SO(5)$, and at first
order in the noncommutative parameter matrix $h\theta^{\mu\nu}$,
we construct smooth monopole --and, some two-monopole-- fields
that solve  the noncommutative Yang-Mills-Higgs equations in the
BPS limit and that are formal power series in $h\theta^{\mu\nu}$.
We show that there exist noncommutative BPS (multi-)monopole
field configurations that are formal power series in
$h\theta^{\mu\nu}$  if, and only if, two a priori free parameters
of the  Seiberg-Witten map take very specific values. These
parameters, that are not associated to field redefinitions nor to
gauge transformations, have thus values that give rise to sharp
physical effects.

\par }
\end{center}

\vspace{20pt} \noindent
 \vfill
\end{titlepage}


\setcounter{page}{2}

\section{Introduction}

Although they have not been detected at the laboratory yet,
monopoles play a key role in the understanding of some properties
of non-abelian gauge theories. In QCD, where monopole degrees of
freedom are uncovered by means of the Abelian projection,  the
confinement of colour can be explained as the effect of monopole
condensation in the vacuum~\cite{'tHooft:1999au}. Monopoles,
namely, BPS monopoles, occur as single-particle states in quantum
non-abelian gauge theories with extended supersymmetry (see
ref.~\cite{Harvey:1996ur} and references therein). S-duality --the
generalization of the Montonen-Olive electric-magnetic
duality conjecture-- seems to be realized in $N=4$
super-Yang-Mills theory and some $N=2$ supersymmetric gauge
theories with vanishing $\beta$-function (for further information
the reader is referred to refs.~\cite{Alvarez-Gaume:1996mv,
Weinberg:2006rq}).

BPS monopoles have been constructed and studied for some
noncommutative $U(N)$ gauge theories~\cite{Hashimoto:1999zw, Bak:1999id,
Hashimoto:1999az, Goto:2000zj, Hashimoto:2000mt, Gross:2000wc, Lechtenfeld:2003vv}.
In particular, in refs.~\cite{Bak:1999id}
and~\cite{Goto:2000zj}, a noncommutative $U(2)$ BPS monopole was
explicitly constructed up to second order in the noncommutative
parameters $\theta^{\mu\nu}$ by expanding the BPS equations in
powers of these parameters. The monopole so obtained is smooth and
goes to the ordinary  $SU(2)$   BPS monopole as
$\theta^{\mu\nu}\rightarrow 0$. And yet, up to the best of our
knowledge, no results concerning the existence and no explicit
construction of monopoles are available so far for noncommutative
gauge theories with simple gauge groups such as $SU(N)$ or
$SO(N)$. It is the main purpose of this paper to look for and give
explicit monopole --and some two-monopole-- solutions to the
noncommutative equations of motion for noncommutative
Yang-Mills-Higgs theories in the BPS
limit when the gauge groups are $SU(2)$, $SU(3)$ and $SO(5)$. Let
us next argue why we have chosen $SU(2)$, $SU(3)$ and $SO(5)$ as
gauge groups.

It has long been known~\cite{Weinberg:1979zt} that in ordinary
Yang-Mills-Higgs theories with simple gauge groups and when there
is  maximal symmetry breaking, all magnetically charged BPS
solutions may be regarded as multi-monopole configurations
containing suitable numbers of different types of the so-called
fundamental monopoles. The fundamental monopoles of the theory are
obtained by embedding the $SU(2)$ BPS monopole in the $SU(2)$
subgroups  of the gauge group of the theory furnished by its
simple roots. Hence it seems natural to start out by constructing
monopole solutions for noncommutative gauge theories with gauge
group $SU(2)$. Once this is done we would like to see how things work
for larger simple gauge groups. The simplest choice seems to
be $SU(3)$. Next, when the gauge symmetry is not broken to the maximal
torus of the gauge group, but the unbroken gauge group has a
non-Abelian component, there exist degrees of freedom that show
the presence of massless monopoles~\cite{Lee:1996vz}. These
massless monopoles do not occur classically as isolated solutions
to the BPS equations and must be studied as  part of
multi-monopole configurations. The simplest instance of a theory
where the existence of these massless monopoles can be analysed was
furnished in ref.~\cite{Lee:1996vz}: it is a theory with gauge
group $SO(5)$ broken down to $SU(2)\times U(1)$.

To formulate a noncommutative field theory whose  gauge group is
$SU(N)$, there is only one available framework. This is the
formalism put forward in refs.~\cite{Madore:2000en, Jurco:2001rq}
that led to the formulation of the noncommutative standard
model~\cite{Calmet:2001na} and some Grand Unified
theories~\cite{Aschieri:2002mc}. The
phenomenology~\cite{Melic:2005su, MohammadiNajafabadi:2006iu,
Alboteanu:2006hh} that these theories gives rise to may be detected at the LHC.

 In the formalism of refs.~\cite{Madore:2000en, Jurco:2001rq}
 --that can be used for any representation of any gauge group-- the
 noncommutative gauge fields are defined from the ordinary fields
 by means of the Seiberg-Witten map, this map being given by a formal
 power series in $\theta^{\mu\nu}$. The noncommutative gauge fields
 thus take values in the enveloping algebra of the Lie algebra of the
 gauge group. This is very much at variance with the standard formalism
used in noncommutative gauge theory, which demands the gauge group
to be $U(N)$. Hence, unlike in the ordinary Minkowski space-time
case, the noncommutative Yang-Mills-Higgs theories to be
considered in this paper are not theories that are part of the
$U(N)$ theories analysed in refs.~\cite{Hashimoto:1999zw,
Bak:1999id, Hashimoto:1999az, Goto:2000zj, Hashimoto:2000mt,
Gross:2000wc}

The layout of this paper is as follows. In section 2 we define our
noncommutative Yang-Mills-Higgs theories and the asymptotic
behaviour of the fields. We also discuss the Bogomol'nyi bound and
deduce the noncommutative BPS equations. The computation of the
most general monopole solution --when it exists-- to the
noncommutative $SU(2)$ BPS equations at first order in
$\theta^{\mu\nu}$ is carried out in section 3. In this section, we
also discuss the existence
of noncommutative fundamental BPS monopoles and some two-monopoles for $SU(3)$ and,
finally, the existence of solutions to the noncommutative BPS
equations that correspond to the family of solutions with
a massless monopole reported in ref.~\cite{Lee:1996vz}
for $SO(5)$. Since, in general, the noncommutative
BPS equations studied in section 3 have no solutions that are
formal power series in $\theta^{\mu\nu}$, we compute in section 4
the static solutions to the  noncommutative Yang-Mills-Higgs
equations with vanishing Higgs potential which go to the ordinary
BPS monopole solutions for $SU(2)$ and to the fundamental and two-monopoles considered previously for$SU(3)$. The computations are
carried out in the gauge $a_0=0$. How the
noncommutative character of space-time affects at first order in
$\theta^{\mu\nu}$ the $SO(5)$ family of solutions with massless
monopoles displayed in ref.~\cite{Lee:1996vz}
is also studied here. In the appendix,  we discuss
whether or not Derrick's theorem implies --as does in the
instanton case, see ref.~\cite{Martin:2005vr}-- that there are no
solutions at second order in $\theta^{\mu\nu}$ to the
noncommutative Yang-Mills-Higgs equations solved in section 4.

\section{The noncommutative Hamiltonian, Bogomol'nyi bounds and the noncommutative BPS equations}

    Our noncommutative gauge theories will have the following action
\begin{equation}
S=\idx -\frac{1}{2}\Tr F_{\m\n}\star F^{\m\n}+\Tr (D_\mu
\Phi)^\dagger\star D^\m \Phi . \label{S}
\end{equation}

The symbol $\star$ will stand for the Moyal product: $(f\star
g)(x) =
f(x)e^{\frac{i}{2}\,h\,\theta^{\mu\nu}\overleftarrow{\partial_{\mu}}\overrightarrow{\partial_{\nu}}}g(x)$.
The noncommutative field strength $F_{\mu\nu}$ and the covariant
derivative $D_{\mu}$ are given by
$F_{\mu\nu}=\partial_{\mu}A_\nu-\partial_{\nu}A_\mu-i[A_\mu,A_\nu]_{\ST}$,
$D_{\mu}=\partial_{\mu}-i[A_{\mu},\phantom{A_{\mu}}]_{\star}$,
respectively. $A_\mu$ and $\Phi$ denote the noncommutative gauge
field and the Higgs field, respectively. They are defined in terms
of the ordinary gauge field, $a_\mu$, and the ordinary Higgs
field, $\phi$, by means of the Seiberg-Witten map, which we shall
take to be  a formal power series in $h\theta^{\mu\nu}$. The
ordinary fields $a_\mu$ and $\phi$ take values in the Lie algebra
of the gauge group --in our case, $SU(2)$, $SU(3)$ and $SO(5)$--
in the fundamental representation. We shall normalize the
generators of the gauge group, the hermitian matrices $T^a$, as
follows $\Tr\, T^a T^b=\frac{1}{2}\,\delta^{ab}$, and assume that
there is a dimensionful parameter $v$ in the theory defined by
$v=2\Tr\,\phi^2(t,|\vec{x}|\rightarrow\infty)$.

The Seiberg-Witten map is not unique --a fact very much welcomed when proving
renormalizability of some models~\cite{Buric:2005xe,Buric:2006wm}. At first
order in $h\theta^{\mu\nu}$ the most general expression
for it that yields
hermitian noncommutative fields and is a polynomial in the fields, their
derivatives and $v$ --we want the map to be well-defined when $v$ vanishes--
reads
\begin{equation}
\begin{array}{l}
A_\mu=\am-\frac{h}{4}\theta^{\alpha\beta}\{a_\alpha,\partial_\beta\am+f_{\beta\m}\}+h D_\mu H+hS_\mu+O(h^2),\\
\Phi=\phi-\frac{h}{4}\theta^{\alpha\beta}\{a_\alpha,2D_\beta\phi+i[a_\beta,\phi]\}+ih[H,\phi]+h F+O(h^2),\\
H=\mu_1\, \theta^{\alpha\beta}f_{\alpha\beta}+\m_2\,\theta^{\alpha\beta}[a_\alpha,a_\beta],\\
S_\mu=\kappa_1\,\theta^{\alpha\beta}D_\m
f_{\alpha\beta}+\kappa_2\,{\theta_\mu}^\beta
\{D_\beta\phi,\phi\}+i
\kappa_3{\theta_\m}^\beta\,[ D_\beta \phi,\phi]+k_4\,v\,{\theta_\mu}^{\beta}D_\beta\phi+
w \,{\theta_\mu}^{\rho}D^{\nu}f_{\nu\rho},\\
F=\lambda_1\theta^{\alpha\beta}\,\{f_{\alpha\beta},\phi\}+i\lambda_2\,\theta^{\alpha\beta}[
f_{\alpha\beta},\phi]+\lambda_3\,v\,\theta^{\alpha\beta}f_{\alpha\beta}.
\end{array}
\label{SWgen}
\end{equation}
The symbols $\mu_i,\,\kappa_i,\, \lambda_i$  and $w$ denote
dimensionless real constants, $v$ is the parameter with mass
dimension defined above, $f_{\mu\nu}=\partial_\mu
a_\nu-\partial_\nu a_\mu-i[a_\mu,a_\nu]$ and
$D_{\mu}=\partial_{\mu}-i[a_\mu,\phantom{a_\mu}]$. When all the
constants $\mu_i,\,\kappa_i,\, \lambda_i$  and $w$ are set to
zero, one gets the standard Seiberg-Witten map, i.e., the
straightforward generalization to our case of the map originally
introduced by Seiberg and Witten for $U(1)$ noncommutative gauge
theories. Notice that the monomials
$\kappa_1\,\theta^{\alpha\beta}D_\m f_{\alpha\beta}$,
$\kappa_3{\theta_\m}^\beta\,[ D_\beta \phi,\phi]$, $w
\,{\theta_\mu}^{\rho}D^{\nu}f_{\nu\rho}$,
$i\lambda_2\,\theta^{\alpha\beta}[ f_{\alpha\beta},\phi]$,
$\kappa_4\,v\,{\theta_\mu}^{\beta}D_\beta\phi$  and
$\lambda_3\,v\,\theta^{\alpha\beta}f_{\alpha\beta}$ always belong
to the Lie algebra of the simple gauge group and thus can be set
to zero by redefining the field $a_\mu$. However, the terms
$\kappa_2\,{\theta_\mu}^\beta \{D_\beta\phi,\phi\}$ and
$\lambda_1\theta^{\alpha\beta}\,\{f_{\alpha\beta},\phi\}$ do not
belong to the Lie algebra of the simple gauge group and hence they
do not correspond to field redefinitions of $a_\mu$. The terms in
eq.~\eqref{SWgen} that go with $H$ are gauge transformations.
Notice that in the noncommutative $U(N)$ case of
refs.~\cite{Hashimoto:1999zw, Bak:1999id, Hashimoto:1999az,
Goto:2000zj, Hashimoto:2000mt, Gross:2000wc, Lechtenfeld:2003vv}
the terms $\kappa_2\,{\theta_\mu}^\beta \{D_\beta\phi,\phi\}$ and
$\lambda_1\theta^{\alpha\beta}\,\{f_{\alpha\beta},\phi\}$ also
correspond to field redefinitions of $a_\mu$. We shall see in the
next section that at least for $SU(2)$, $SU(3)$ and $SO(5)$, and at odds
with the $U(N)$ case, the value of the real constants $\kappa_2$
and $\lambda_1$ is physically relevant.

In this paper we will not be interested in the the most general
Seiberg-Witten map. Indeed, in keeping with the situation for the
noncommutative $U(N)$ theories of refs.~\cite{Hashimoto:1999zw,
Bak:1999id, Hashimoto:1999az, Goto:2000zj, Hashimoto:2000mt,
Gross:2000wc}, we shall restrict ourselves to theories whose
action in the temporal gauge --here, $a_0=0$-- depends on the
generalized coordinates --$a_i(t,\vec{x})$ and $\phi(t,\vec{x})$,
in our case--, the generalized velocities --$\partial_0
a_i(t,\vec{x})$ and $\partial_0\phi(t,\vec{x})$, for our
theories-- and the spatial derivatives of them, but not on
generalized accelerations nor on any other higher time
derivatives. Thus, the noncommutative matrix parameter
$\theta^{\m\n}$ will be taken to be of magnetic type --i.e.,
$\theta^{0i}=0$-- and $\Phi[\phi,a_\mu]$ and $A_i[a_\mu,\phi]$
must not involve time derivatives --otherwise $D_0\Phi$ or
$F_{0i}$ in eq.~\eqref{S} would give rise, at least, to second order time
derivatives. $a_0=0$ does not imply $A_0=0$, but
restricts the form of $A_0$ to linear combinations of terms linear
in $(\partial_0 a_i, \partial_0\phi)$, the coefficients of these
combinations being functions of the ordinary fields
and/or  their spatial derivatives,
but having no time derivatives of the former. For this
Seiberg-Witten map, $F_{ij}$ and $D_i\Phi$ do not involve time
derivatives of ordinary fields, and $F_{0i}$ and $D_0\Phi$ are
linear combinations of terms proportional to $\partial_0 a_i, \partial_0\phi$
, with coefficients free of time derivatives. That a
Seiberg-Witten map --in fact, infinitely many-- satisfying the
previous requirements exists at any order in $h\theta^{\m\n}$ can
be readily shown by using the Seiberg-Witten map defined by the
following equations:
\begin{equation}
\begin{array}{l}
    \frac{dA_\mu}{dh}=-\frac{1}{4}\theta^{ij}\{A_i,\partial_j A_\mu+F_{j\mu}\}_\star+D_{\mu}\hat{H}+\hat{S}_\m\\\
    \frac{d\Phi}{dh}=-\frac{1}{4}\theta^{ij}\{A_i,2 D_j \Phi+i[A_j,\Phi]_\star\}_\star+i[\hat{H},\Phi]_{\star}+{\hat F}\\
\hat{H}=\mu_1\,\theta^{ij}F_{ij}+\m_2\,\theta^{ij}[A_i,A_j]_{\star},\\
\hat{S}_\mu=\kappa_1\,\theta^{i j}D_\m F_{i
j}+\kappa_2\,{\theta_\mu}^j \{D_j\Phi,\Phi\}_{\star}+i
\kappa_3{\theta_\m}^j\,[ D_j \Phi,\Phi]_{\star}+\kappa_4\,v\,{\theta_\m}^j D_j\Phi,\\
\hat{F}=\lambda_1\theta^{ij}\,\{F_{ij},\Phi\}+i\lambda_2\,\theta^{ij}[
F_{ij},\Phi]_{\star}+\lambda_3\,v\,\theta^{ij}F_{ij},
\end{array}
\label{Swanyord}
\end{equation}
where $\mu_i,\,\kappa_i$ and $\lambda_i$  are dimensionless real
constants, $F_{\mu\nu}=\partial_\mu A_\nu-\partial_\nu
A_\mu-i[A_\mu,A_\nu]_{\star}$ and
$D_{\mu}=\partial_{\mu}-i[A_\mu,\phantom{A_\mu}]_\star$.

The restrictions imposed on the Seiberg-Witten map in the previous
paragraph do not give a unique Seiberg-Witten map, though. At first order
in $h\theta^{\mu\nu}$, they merely set $w=0$. However, this yields
an action that is quadratic in the generalized velocities so that
the Hamiltonian can be derived from it by using the
standard textbook formalism. Indeed, there is a single
generalized momenta, $p_i=\frac{\partial {\cal L}}{\partial
\dot{q}_i}$, per generalized coordinate, $q_i$, and the
Hamiltonian, ${\cal H}$, can be obtained from the Lagrangian,
${\cal L}$, by employing the elementary expression ${\cal
H}=\sum_i\, p_i\dot{q}_i-{\cal L}$. In our case the Hamiltonian
reads
\begin{equation}
    {\cal H}=\id3x\;\Tr\left(E_{i}E_{i}+B_{i}B_{i}+D_0\Phi D_0\Phi+D_i\Phi D_i \Phi\right),
\label{Hcan}
\end{equation}
where $E_i=F_{i0},\,B_i=\frac{1}{2}\epsilon_{ijk}F_{jk}$. The
Hamiltonian has been computed in the gauge $a_0=0$;
 the Gauss-law constraint takes here the form
\begin{equation}
\Tr\frac{\delta A_0}{\delta a^a_0}\left(D_j
E_j+i[D_0\Phi,\Phi]_\star\right)=0. \label{Gauss}
\end{equation}
Let us note that although the Hamiltonian in eq.~\eqref{Hcan} is
defined by the same formal expression as in the $U(N)$ case of
refs.~\cite{Hashimoto:1999zw, Bak:1999id,
Hashimoto:1999az, Goto:2000zj, Hashimoto:2000mt, Gross:2000wc}, the Gauss-Law
constraint signals a difference with the $U(N)$ case, where it
reads $D_j E_j+i[D_0\Phi,\Phi]_\star =0$. This difference stems from
the  fact that for simple gauge groups, unlike for $U(N)$
gauge groups, noncommutative fields do not take values in the Lie
algebra of the gauge group.

We shall introduce next  the asymptotic boundary conditions for
the noncommutative fields $\Phi(t,\vec{x})$ and
$A_\mu(t,\vec{x})$. These conditions read
\begin{equation}
\begin{array}{l}
\Phi(t,\vec{x})\;\sim\;\phi(t,\vec{x})\;+\;O(\frac{1}{|\vec{x}|^2})\quad\quad\quad\text{as}\quad\quad\quad
|\vec{x}|\rightarrow\infty,\\
A_\mu (t,\vec{x})\;\sim\; a_\mu
(t,\vec{x})\;+\;O(\frac{1}{|\vec{x}|^2})\quad\quad\quad\text{as}\quad\quad\quad
|\vec{x}|\rightarrow\infty.\\
\end{array}
\label{noncboundary}
\end{equation}
A simple dimensional analysis shows that the asymptotic boundary
conditions above follow from the Seiberg-Witten map defined at
first order by eq.~\eqref{SWgen} --and at higher-order  by
eq.~\eqref{Swanyord}-- and the asymptotic boundary conditions for
the ordinary fields  that we set next. For the ordinary fields
$\phi(t,\vec{x})$ and $a_i(t,\vec{x})$, we shall take the boundary
conditions in the gauge $a_0=0$ that are customary in monopole
physics~\cite{Coleman:1975qj}:
\begin{equation}
\begin{array}{l}
\phi(t,\vec{x})\;=\;g(t,\hat{x})\phi_0
g(t,\hat{x})^{\dagger}\;+\;O(\frac{1}{|\vec{x}|})\quad\quad\quad\text{as}\quad\quad\quad
|\vec{x}|\rightarrow\infty,\\
a_i (t,\vec{x})\;\sim\;
\frac{1}{|\vec{x}|}\quad\quad\quad\text{as}\quad\quad\quad
|\vec{x}|\rightarrow\infty,\\
D_i\phi\;\sim\;
\frac{1}{|\vec{x}|^2}\quad\quad\quad\text{as}\quad\quad\quad
|\vec{x}|\rightarrow\infty,\\
\end{array}
 \label{ordboundary}
 \end{equation}
 where $\hat{x}=\vec{x}/|\vec{x}|$ and $\phi_0$ is the value of the
 Higgs field along  a given fixed direction in space.
 $g(t,\hat{x})$ defines a smooth map from the two-sphere at spatial infinity into
the coset $G/H$, $G$ and $H$ being respectively the broken and
unbroken gauge groups.

Let us introduce now the magnetic charge, $Q_M$, of the
noncommutative fields:
\begin{equation}
Q_M\;=\;\frac{1}{2\pi v}\Tr\int\!dS_i\; B_i \Phi\;=\;\frac{1}{2\pi
v}\Tr\int\!dS_i\; b_i \phi . \label{ncmagcharge}
\end{equation}
$b_i$ and $\phi$ are the ordinary field configurations that yield
$B_i$ and $\Phi$ upon acting with the Seiberg-Witten map. The
integrals are carried out over a two-sphere at spatial infinity
and
$v=(2\,\Tr\,\phi^2(|\vec{x}|\rightarrow\infty))^{\frac{1}{2}}=(2\,\Tr\,\phi_0^2)^{\frac{1}{2}}$.
$Q_M$ depends only on the boundary conditions for the fields. The
equality between the two surface integrals above follows from the
asymptotic boundary conditions in eq.~\eqref{noncboundary} and in
turn  implies that the noncommutative fields carry the same
magnetic charges as the BPS (multi-)monopoles of the corresponding
ordinary theory. Indeed, both the boundary conditions in
eq.~\eqref{ordboundary} and the form of the Seiberg-Witten map in
eqs.~\eqref{SWgen} and~\eqref{Swanyord} lead to the conclusion
that at very large distances the chief contributions to the
equations of motion of our noncommutative theory are given by the
corresponding ordinary Yang-Mills-Higgs equations. Of course,
$Q_M$ above is constrained by the quantization condition of
ref.~\cite{Goddard:1976qe}.

Let us apply now the Bogomol'nyi trick to the r.h.s. of
eq.~\eqref{Hcan}:
\begin{equation}
    {\cal H}=\idx\Tr\left(D_0\Phi D_0\Phi+E_i E_i+(B_i\mp D_i\Phi)^2\pm 4\pi\, v\, Q_M\right)\geq 4\pi\,v\,|Q_M|.
\label{Hambogo}
\end{equation}
    Hence, for each value of $Q_M$ --as in the ordinary case--, the absolute
minima of the energy are given by the  solutions to the equations
\begin{equation}
    B_i=\pm D_i\Phi,\quad D_0\Phi=0,\quad E_i=0.
\label{BPS1}
\end{equation}
These equations are the noncommutative BPS equations. Notice that
they are the straightforward generalization to
noncommutative space-time of the ordinary BPS equations. Also
notice that the noncommutative BPS equations above imply the
Gauss-law constraint in eq.~(\ref{Gauss}).

That the meaning and form of the noncommutative BPS equations is
analogous to those of the  ordinary BPS equations and that the
magnetic charge of the noncommutative field configurations is the
same as that of their ordinary counterparts are facts that our
theories share in common with the $U(N)$ noncommutative
theories studied in ref.~\cite{Hashimoto:1999zw,
Bak:1999id,Goto:2000zj}. However, we shall see in the next section
that the BPS moduli spaces of our theories are quite different
from the corresponding spaces of the $U(N)$ case.

To close this section let us point out that the solutions to the noncommutative
BPS equations in eq.~\eqref{BPS1} are also solutions to the Yang-Mills-Higgs
equations derived from the action in eq.~\eqref{S}. The latter equations
read
\begin{equation}
\begin{array}{l}
\idx\,\Big\{\Tr\Big[\frac{\delta A_\nu(x)}{\delta a^a_{\mu}(y)}
\big\{D_{\rho}F^{\rho\nu}(x)-i[\Phi, D^{\nu}\Phi]_\star(x)\big\}\Big]-
\Tr\Big[\frac{\delta \Phi(x)}{\delta a^a_{\mu}(y)}
\big\{D_{\rho}D^{\rho}\Phi(x)\big\}\Big]\Big\}=0\\
\idx\,\Big\{\Tr\Big[\frac{\delta A_\nu(x)}{\delta \phi^a(y)}
\big\{D_{\rho}F^{\rho\nu}(x)-i[\Phi, D^{\nu}\Phi]_\star(x)\big\}\Big]-
\Tr\Big[\frac{\delta \Phi(x)}{\delta \phi^a(y)}
\big\{D_{\rho}D^{\rho}\Phi(x)\big\}\Big]\Big\}=0.\\
\label{NCYMH}
\end{array}
\end{equation}


\section{Solutions to the noncommutative BPS equations}

    In this section we  shall  look for solutions to the BPS equations given
in eqs.~(\ref{BPS1}) that are formal power series in
$h\theta^{\mu\nu}$. We shall work in the temporal gauge $a_0=0$
and consider the following  (broken) gauge groups: $SU(2)$, $SU(3)$
and $SO(5)$. These groups will be  broken down to $U(1)$,
$U(1)\times U(1)$ and $SU(2)\times U(1)$, respectively, by
choosing appropriate asymptotic boundary conditions for the Higgs
field.

Let us recall --see previous section--  that our Seiberg-Witten
map --for $a_0=0$-- is such that $A_0$ is linear in
$(\dot{a}^a_i=\partial_0 a^a_i,\dot{\phi}^a=\partial_0\phi^a)$ with
coefficients that are constructed only with  $a_i$, $\phi$ and
$\partial_k$, and that $A_i$ and $\Phi$ only depend on $a_i$,
$\phi$ and their spatial partial derivatives. Then,
\begin{equation*}
\begin{array}{l}
 A_0=\sum_{l>0}h^lL_0^{(l)ia}[\theta^{\mu\nu},a_k,\phi,\partial_k]\dot{a}^a_i+
\sum_{l>0}h^lM_0^{(l)a}[\theta^{\mu\nu},a_k,\phi,\partial_k]\dot{\phi}^a\\
F_{0i}=\dot{a}_i+\sum_{l>0}h^lP^{(l)ja}_{0i}[\theta^{\mu\nu},
a_k,\phi,\partial_k]\dot{a}^a_j+\sum_{l>0}h^lQ^{(l)a}_{0i}[\theta^{\mu\nu},
a_k,\phi,\partial_k]\dot{\phi}^a\\
D_0\Phi=\dot{\phi}+\sum_{l>0}h^lS^{(l)ja}_{0}[\theta^{\mu\nu},
a_k,\phi,\partial_k]\dot{a}^a_j+\sum_{l>0}h^lT^{(l)a}_{0}[\theta^{\mu\nu},
a_k,\phi,\partial_k]\dot{\phi}^a,\\
\end{array}
\end{equation*}
where $L_0^{(l)ia}$, $M_0^{(l)a}$, $P^{(l)ja}_{0i}$, $Q^{(l)a}_{0i}$,
$S^{(l)ja}_{0}$ and $T^{(l)a}_{0}$ are homogeneous polynomials in $\theta^{\mu\nu}$ of degree $l$.
The previous expressions lead to the conclusion that if
$a_i$ and $\phi$ are formal power series in $h\theta^{\mu\nu}$, the following
result holds
\begin{displaymath}
E_i=0\quad\text{and}\quad D_0\Phi=0\quad\Longleftrightarrow \quad\dot{a_i}=0\quad \text{and}\quad \dot{\phi}=0.
\end{displaymath}
Hence, in the remainder of this section, we shall look for
solutions to $B_i=\pm D_i\Phi$ that are time independent and are given by
the following formal expansions in powers of $h\theta^{\mu\nu}$:
\begin{equation}
a_i=a_i^{(0)}+\sum_{l>0}h^la_i^{(l)},\quad\phi=\phi^0+\sum_{l>0}h^l\phi^{(l)}.
\label{expfields}
\end{equation}
$a_i^{(l)}$ and $\phi^{(l)}$ are homogeneous polynomials in $\theta^{\mu\nu}$
of degree $l$. We shall use besides the following power series in $h\theta^{\m\n}$:
\begin{equation}
f_{ij}=f_{ij}^{(0)}+\sum_{l>0}h^lf_{ij}^{(l)},\quad
D_k\phi=(D_k\phi)^0+\sum_{l>0}h^l(D_k\phi)^{(l)},
\label{expstrength}
\end{equation}
where $f_{ij}^{(0)}=\partial_{i}a_j^{(0)}-\partial_{j}a_i^{(0)}-i[a_i^{(0)},a_j^{(0)}]$
and $(D_k\phi)^0=\partial_{k}\phi^{(0)}-i[a_k^{(0)},\phi^{(0)}]$, and $f_{ij}^{(l)}$ and $(D_k\phi)^{(l)}$
are also homogeneous polynomials in $\theta^{\mu\nu}$ of degree $l$.


\subsection{$SU(2)$ noncommutative BPS magnetic (anti-)monopoles}

Let us seek for time-independent $a_i$ and $\phi$  that belong to
the Lie algebra of $SU(2)$ in its fundamental representation and that
solve $B_i=\pm D_i\Phi$ at first order in $h\theta^{\mu\nu}$. We
shall further assume that the asymptotic boundary conditions  are
such that $Q_M=\pm 1$ --see eq.~\eqref{ncmagcharge}, i.e., we
shall look for noncommutative BPS monopoles and anti-monopoles.

 We shall  begin our analysis by assuming that the noncommutative fields
are defined by the standard form of the Seiberg-Witten map. This form
is obtained by setting $\hat{H}, \hat{S}_\mu$ and $\hat{F}$ in
eq.~\eqref{Swanyord} to zero.
    For the standard form of the Seiberg-Witten map in the  gauge
    $a_0=0$ and for time-independent field configurations, it is easy to see that the field
$\Phi$ is defined by the standard form of the Seiberg-Witten  map that
corresponds to the $A_4$ component of the gauge field in a
noncommutative space-time with  Euclidean signature. Hence, we can
combine $a_i$ and $\phi$ into an Euclidean ordinary gauge field
$\overline{a}_\mu=(a_i,a_4=\phi)$  and $A_i$ and $\Phi$ into a
noncommutative gauge field $\overline{A}_\mu=(A_i,A_4=\Phi)$, so
that, again, the standard form of the Seiberg-Witten map defines
$\overline{A}_\mu$ in terms of $\overline{a}_\mu$. Now, with the
definition $\overline{F}_{\mu\nu}=\partial_\mu\overline{A}_\nu
-\partial_\nu\overline{A}_\mu-i[\overline{A}_\mu,\overline{A}_\nu]_\star$
and recalling that neither $a_\mu$, nor $A_\mu$, depend on $x^4$,
one concludes that the BPS equations in eq.~\eqref{BPS1} can be
turned into the following (anti-)self-duality equations:
\begin{equation*}
\overline{F}_{\mu\nu}=\pm \tilde{\overline{F}}_{\mu\nu},\quad
\tilde{\overline{F}}_{\mu\nu}=\frac{1}{2}\epsilon_{\m\n\r\s}\overline{F}_{\r\s}.
\end{equation*}
 Unfortunately, it has been shown in ref.~\cite{Martin:2005vr} that even
 at first order in $h\theta^{\mu\nu}$ there are no solutions
to the previous equation. There are thus no  noncommutative
(anti-)monopoles arising from the noncommutative $SU(2)$ BPS
equations for the standard form of the Seiberg-Witten map. Hence,
all that remains for us to do is to see whether or not this
negative result can be turned into a positive one by taking advantage of
the ambiguities in the form of the Seiberg-Witten map that do not
correspond neither to field redefinitions nor to gauge
transformations.

 For the general form --with $w=0$-- of the Seiberg-Witten map given
 in eq.~\eqref{SWgen}, the previous construction, that turns the BPS
 equations into the (anti-)self-duality equations above, cannot be
 carried out. Hence, we have to deal with the equation $B_i=\pm
 D_i\Phi$ directly. At zero order in $h\theta^{\mu\nu}$, the
 previous equation is the ordinary equation:
\begin{equation}
b_i^{(0)}=\pm (D_i\phi)^{(0)},
\label{BPSord}
\end{equation}
where $b_i^{(0)}=\frac{1}{2}\epsilon_{ijk}f_{jk}^{(0)}$ and
$(D_i\phi)^{(0)}=\partial_i\phi^{(0)}-i[a_i^{(0)},\phi^{(0)}]$.
$a_i^{(0)}$,  $\phi^{(0)}$,  $f_{ij}^{(0)}$ have been defined in  eqs.~\eqref{expfields} and~
\eqref{expstrength}.

The solutions to  eq.~\eqref{BPSord} with magnetic charge $\pm1$ are the
ordinary $SU(2)$ (anti-)monopoles in the fundamental representation:
\begin{equation}
\begin{array}{l}
\phi^{(0)}= \frac{x^a}{r}H(r)\,\frac{\sigma^a}{2},\quad H(r)=\pm(\frac{1}{r}-\lambda\coth \lambda r)\\
a_i^{(0)}=[1-K(r)]\,\epsilon_{ial}\,\frac{x^l}{r^2}\,\frac{\sigma^a}{2},\quad
K(r)=2-\frac{\lambda r}{\sinh \lambda r}.
\end{array}
\label{BPS}
\end{equation}
where $\{\sigma^a\}_{\{a=1,2,3\}}$ stands for the Pauli matrices
and $\lambda=v$ --later on we will consider $SU(2)$ monopoles
embedded in $SU(3)$ and the value of $\lambda$ will change.

The Seiberg-Witten map gives rise to the following expressions for
the noncommutative objects $F_{ij}$ and $D_k\Phi$ defined as power series
in $h\theta^{\mu\nu}$:
\begin{equation}
F_{ij}=f_{ij}+\sum_{l>0}h^l F^{(l)}_{ij}[a_k,\phi],\quad
D_k\Phi=D_k\phi+\sum_{l>0}h^l {\cal O}_k^{(l)}[\phi,a_i].
\label{thetaexp1}
\end{equation}
Since  $a_i$ and $\phi$ are defined by the expansions in
eq.~(\ref{expfields}), we end up with the following results
\begin{equation}
\begin{array}{l}
F^{(l)}_{ij}[a_k,\phi]=\sum_{m\geq 0}h^m F_{ij}^{(l,\,m)},\quad F^{(l,\,m)}_{ij}=
\frac{1}{m!}\frac{d^m}{dh^m}F_{ij}^{(l)}[a_k,\phi]|_{h=0}\\
{\cal O }_k^{(l)}[\phi,a_k]=\sum_{m\geq 0}h^m {\cal
O}_k^{(l,\,m)},\quad {\cal
O}_k^{(l,\,m)}=\frac{1}{m!}\frac{d^m}{dh^m}{\cal
O}_k^{(l)}[\phi,a_k]|_{h=0}.
\end{array}
\label{thetaexp2}
\end{equation}
 We are now ready to write down the contribution to $B_i=\pm D_i\Phi$ that is of order one in
 $h\theta^{\m\n}$:
\begin{equation}
(f_{ij}^{(1)}+F_{ij}^{(1,0)})=\pm\epsilon_{ijk}[(D_k\phi)^{(1)}+{\cal O}_k^{(1,0)}].
\label{monopoleh}
\end{equation}
The objects that occur in this equation have been defined in
eqs.~\eqref{expfields}, ~\eqref{expstrength},  ~\eqref{thetaexp1} and ~\eqref{thetaexp2}.

Both sides of eq.~\eqref{monopoleh} take values in the universal
enveloping algebra of $SU(2)$ in the fundamental representation.
Hence, both sides of eq.~\eqref{monopoleh} can be expressed as a
linear combination of the $2\times 2$ identity matrix and the
Pauli matrices. We thus conclude that eq.~\eqref{monopoleh} is
equivalent to the set of equations $a)$ and $b)$ that follow:
\begin{equation}
\begin{array}{l}
a)\quad\Tr\big[(f_{ij}^{(1)}+F_{ij}^{(1,0)})\big]=\pm\epsilon_{ijk}\Tr\big[(D_k\phi)^{(1)}+{\cal
O}_k^{(1,0)}\big],\\
b)\quad\Tr\big[\frac{\sigma^a}{2}(f_{ij}^{(1)}+F_{ij}^{(1,0)})\big]=\pm\epsilon_{ijk}
\Tr\big[\frac{\sigma^a}{2}\big((D_k\phi)^{(1)}+{\cal
O}_k^{(1,0)}\big)\big].\\
\end{array}
\label{idenproj}
\end{equation}
Some little algebra leads to the result that $a)$ in
eq.~\eqref{idenproj} is equivalent to
\begin{equation}
\begin{array}{l}
\sum_a\frac{1}{2}[(f^{(0),\, a}_{12})^2+(f^{(0),\, a}_{13})^2+(f^{(0),\, a}_{23})^2]
\theta_{ij}=\kappa_2{\theta_{jk}}
\,\partial_i(\partial_k\phi^{(0),\, a}\phi^{(0),\, a})-(i\leftrightarrow j)\\
\phantom{\sum_a\frac{1}{2}[(f^{(0),\, a}_{12})^2+(f^{(0),\,
a}_{13})^2+(f^{(0),\, a}_{23})^2]\theta_{ij}=}
\pm\lambda_1\epsilon_{ijk}\,\partial_k\,[\,\theta^{mn}f_{mn}^{(0),\,
a}\phi^{(0),\, a}].
\end{array}
\label{consist}
\end{equation}
Since $a_i^{(0)}=a_i^{(0),\,a}\,\frac{\sigma^a}{2}$ and
$\phi^{(0)}=\phi^{(0),\,a}\,\frac{\sigma^a}{2}$ are fixed by
eq.~\eqref{BPS}, one concludes that eq.~\eqref{consist} --and hence $a)$ in
eq.~\eqref{idenproj}-- is more a no-go condition than an equation
of motion. Indeed, it holds if, and only if, the parameters
$\kappa_2$ and $\lambda_1$ of the Seiberg-Witten map in
eq.~\eqref{SWgen} are tuned to the following values
\begin{equation}
\kappa_2=-\frac{1}{2},\quad\lambda_1=\frac{1}{4}.
\label{solconsist}
\end{equation}

Next, taking into account the Seiberg-Witten map  defined in
eq.~\eqref{SWgen}, one may show that the equality $b)$ in
eq.~\eqref{idenproj} holds if, and only if,
\begin{equation}
D^{(0)}_i(a'_j)-D^{(0)}_j(a'_i)=\pm\epsilon_{ijk}\,\big(D^{(0)}_k\phi'-i[a'_k,\phi^{(0)}]\big),
\label{zeromodeeq}
\end{equation}
where
\begin{equation}
\begin{array}{l}
a'_i=a^{(1)}_i+\kappa_1\,\theta^{kl}\,D^{(0)}_i f^{(0)}_{kl}+i\kappa_3\,{\theta_i}^l\,
[(D_l\phi)^{(0)},\phi^{(0)}]+\kappa_4\,v{\theta_i}^j D^{(0)}_j\phi^{(0)},\\
\phi'=\phi^{(1)}+i\lambda_2\,\theta^{kl}\,[f_{kl}^{(0)},\phi^{(0)}]+\lambda_3\,v\,\theta^{ij}f_{ij}^{(0)}.
\end{array}
\label{defone}
\end{equation}
Now, eq.~\eqref{zeromodeeq} is the equation of the zero modes
associated to the ordinary $SU(2)$ BPS (anti-)monopole. Hence,
$a'_i, \phi'$ in eq.~\eqref{defone} satisfy the zero mode
equations in the background of the ordinary $SU(2)$ BPS
(anti-)monopole. Also notice that eq.~\eqref{zeromodeeq} shows
that the monomials $\kappa_2\,\theta_i^{\;j}\{D_j\phi,\phi\}$ and
$\lambda_1\, \theta^{ij}\{f_{ij},\phi\}$ do not contribute to $b)$
in eq.~\eqref{idenproj}, so that the latter  equations are not
affected by the constraint in eq.~\eqref{solconsist}.

Let us stress that  at first
order in $h\theta^{\mu\nu}$ only for the choice of constants given
in eq.~\eqref{solconsist} there exist BPS (anti-)monopole
solutions to the noncommutative BPS equations defined with the
help of the Seiberg-Witten map --with $w=0$-- in
eq.~\eqref{SWgen}. These solutions are given by the ordinary
(anti-)monopoles plus the field redefinitions that the terms of
the Seiberg-Witten map which go with $\kappa_1$, $\kappa_3$ and
$\lambda_2$ give rise to. From the previous statement one deduces
that the terms  $\kappa_2\,\theta_i^{\;j}\{D_j\phi,\phi\}$ and
$\lambda_1\, \theta^{ij}\{f_{ij},\phi\}$ in eq.~\eqref{SWgen} that
constitute part of the ambiguity in the Seiberg-Witten map --the other
being field redefinitions and gauge transformations-- are not
physically irrelevant in the $SU(2)$ case since the existence of a
BPS moduli space with elements that are formal power series in
$h\theta^{\m\n}$ depends drastically on the value of $\kappa_2$
and $\lambda_1$.

We shall close this subsection showing that the number of zero
modes, or moduli, associated with the noncommutative BPS monopole
found is four. Indeed, the noncommutative BPS equations are
invariant under translations --three moduli--and the large gauge
transformation $e^{i\chi\frac{\phi(\vec{x})}{v}}$, $0\leq \chi <
2\pi$ --one moduli. One may rule out the possibility of the
existence of further zero modes --that should vanish as
$\theta^{\mu\nu}\rightarrow 0$-- as follows. Let  $\delta
z=(\delta a_i,\delta\phi)$ denote a zero mode that can be
expressed as a power series in $h\theta^{\mu\nu}$: $\delta z
=\sum_{l \geq 0}\, h^l\delta z^{(l)}$. Then the components of $\delta z^{(l)}$, which are homogeneous polynomials in
$\theta^{\m\n}$, must satisfy the following system
of equations
\begin{equation*}
L^{(0)}\delta z^{(0)}=0,\quad
L^{(0)}\delta z^{(l)}=f^{(l)}[a_i^{(m)},\phi^{(p)},\delta z^{(q)}],
\end{equation*}
where $L^{(0)}$ is the ordinary operator characterizing the ordinary zero
modes:
\begin{equation*}
 (L^{(0)}\delta z)_i =\epsilon_{ijk}D_j \delta a_k\mp(D_i\delta \phi-i[\delta a_i,\phi])
\end{equation*}
and $f^{(l)}$ is a homogeneous polynomial of degree $l$ in
$\theta^{\mu\n}$. The actual value of $f^{(l)}$ is immaterial to our argument,
but for the fact that it does not depend on $\delta z^{(l)}$. Now,
let us assume that there exists a solution to the previous set of
equations; then, there are as many solutions as there are
solutions to $L^{(0)}\delta z=0$. We know that, modulo gauge
transformations that go to the identity at infinity, the number of
linearly independent solutions to the ordinary zero mode equation
is four.

\subsection{Fundamental noncommutative BPS monopole configurations for $SU(3)$. Two-monopole configurations}

In this subsection the ordinary fields $a_i$ and $\phi$ in eq.~\eqref{expfields} will
take values in the Lie algebra of $SU(3)$ in the fundamental representation. Let us further
assume that the asymptotic value of $\phi$  --and, hence, the asymptotic value of
$\Phi$, see eq.~\eqref{noncboundary}--  along the negative $z$-axis is given by
\begin{equation}
\phi(0,0,z\rightarrow-\infty)=v\,\vec{h}\cdot\vec{H},
\label{hsu3}
\end{equation}
where $\vec{H}=(H_1,H_2)$, $H_1$ and $H_2$ being the generators of
the Cartan subalgebra of $SU(3)$, and $\vec{h}=(h_1,h_2)$ is a
unitary two-vector that unless otherwise stated will  have
non-vanishing scalar product with any root of $SU(3)$. 

For these
boundary conditions the gauge $SU(3)$ symmetry  is broken down to
$U(1)\times U(1)$. It is  well known~\cite{Goddard:1976qe} that
for this maximal breaking a solution to the ordinary BPS equations, $b_i=D_i\phi$,
will have a magnetic vector $\vec{g}=n_1
\frac{\vec{\beta}_1}{\beta_1^2} + n_2
\frac{\vec{\beta}_2}{\beta_2^2}$, where the integers $n_1\geq 0$ and
$n_2\geq 0$  are topological numbers
and $\vec{\beta}_1$ and $\vec{\beta}_2$ are the unique set of simple roots of
$SU(3)$ selected by the condition $\vec{h}\cdot \beta_{i}>0$.
It
is further well established~\cite{Weinberg:1979zt} that these BPS solutions can
be understood as multi-monopole configurations made out of two
fundamental monopole solutions or their corresponding anti-monopoles. These fundamental monopole
solutions have topological charges $(n_1,n_2)$ equal to $(1,0)$
and $(0,1)$, respectively, and are obtained by embedding the
$SU(2)$ monopole in the $SU(2)$ subgroups of $SU(3)$ defined by
the roots$\vec{\beta}_1$ and $\vec{\beta}_2$ of $SU(3)$, respectively.

Let $T^{a}_{\beta_i}$, $a=1,2,3$ and $i=1,2$ be the generators of the $SU(2)$
subgroup of $SU(3)$ defined by the simple root $\vec{\beta}_i$. Then,
\begin{equation*}
    T_{\beta_i}^1=\frac{1}{\sqrt{2 \beta_i^2}}\,(E_{\vec{\beta}_i}+
E_{-\vec{\beta}_i}),\quad T_{\beta_i}^2=\frac{-i}{\sqrt{2 \beta_i^2}}\,
(E_{\vec{\beta}_i}-E_{-\vec{\beta}_i}),\quad T_{\beta_i}^3=
\frac{1}{\beta_i^2}\,\vec{\beta}_i\cdot\vec{H},
\end{equation*}
where $E_{\vec{\beta}_i}$ is the generator of $SU(3)$ defined by the
root $\vec{\beta}_i$ in the Cartan-Weyl decomposition of the Lie algebra of $SU(3)$:
$[H_k,E_{\vec{\beta}_i}]=(\vec{\beta}_i)_k\,E_{\vec{\beta}_i}$. The fundamental
monopoles with topological charges $(1,0)$ and $(0,1)$ are obtained by replacing
$i$ with $1$ and $2$, respectively, in the following equations
\begin{equation}
\begin{array}{l}
\phi^{(0)}_{\beta_i}=\sum_{a=1,2,3}\;\phi^{(0)\,a}\,T_{\beta_i}^a+v
\vec{h}\cdot\vec{H}-v \vec{h}\cdot\vec{\beta}_i\,T_{\beta_i}^3\\
a_{\beta_i}^{(0)}=\sum_{a=1,2,3}\; a_i^{(0)\,a}\,T^a_{\beta_i}.
\end{array}
\label{order0}
\end{equation}
$\phi^{(0)\,a}$ and $a_i^{(0)\,a}$ are the functions given in eq.~(\ref{BPS}) with the choice of positive sign for $H(r)$ and
for $\lambda=v\,\vec{h}\cdot\vec{\beta}_i$. Of course, the previous field configurations
are solutions to the noncommutative BPS equations $B_i=  D_i\Phi$ at order
zero in $h\theta^{\mu\nu}$.

Before computing the first-order-in-$\theta^{\m\n}$ corrections
--$a_i^{(1)}$ and $\phi^{(1)}$ in eq.~\eqref{expfields}-- to the
previous ordinary fundamental monopoles, we need some
preparations. We shall choose the coordinate axis in the root
space and the Cartan-Killing metric so that $\vec{\beta}_1=(1,0)$
and $\vec{\beta}_2=(-\frac{1}{2},\frac{\sqrt{3}}{2})$. The
Gell-Mann generators of $SU(3)$ will be denoted by
$T^a=\frac{\lambda^a}{2}$, $a=1\dots 8$, where $\lambda^a$ are the
Gell-Mann matrices. Under the adjoint action of the $SU(2)$
generators $T^a_{\beta_i}$, $a=1,2,3$, the generators of $SU(3)$
$T^a$, $a=1\dots 8$ can be sorted out into one triplet,
two doublets and one singlet, which have the following value in
terms of Gell-Mann matrices,
\begin{equation}
\begin{array}{l}
\begin{array}{lll}
\beta_1:\, \text{Triplet:}\,\,\{T^1,T^2,T^3\},&\text{Doublets:}\,
\{T^4,T^5,T^6,T^7\},&\text{Singlet:}\,\,T^8,\\
\beta_2:\, \text{Triplet:}\,\{T^6,T^7,-\frac{1}{2}T^3+\frac{\sqrt{3}}{2}T^8\},
&\text{Doublets:}\,\{T^1,T^2,T^4,T^5\},&\text{Singlet:}-\frac{\sqrt{3}}{2}T^3-\frac{1}{2}T^8.\\
\end{array}
\end{array}
\label{bases}
\end{equation}
     Denoting by $T_\beta^s$ the singlet generator in the previous equation,
it can be seen that the ordinary field configurations of eq.~(\ref{order0})
can be written thus
\begin{equation}
    \phi^{(0)}=\sum_{\text{triplet}}\phi^{t\,a} T^a_\beta+\phi^s T^s_\beta,
\quad\phi^s=2\,v\,\Tr\,(T^s_\beta\, \vec{h}\cdot\vec{H}).
\label{embed}
\end{equation}
$\phi^{t\,a}$, $a=1,2,3$, are given by the components of the ordinary
$SU(2)$ monopole.\par

We are now ready to compute $a^{(1)}_i$ and $\phi^{(1)}$ in
eq.~\eqref{expfields} in the case at hand. Proceeding as in the
$SU(2)$ case --see eqs.~\eqref{thetaexp1} and~\eqref{thetaexp2}--,
one obtains that $a^{(1)}_i$ and $\phi^{(1)}$ must satisfy the
following equation:
\begin{equation}
(f_{ij}^{(1)}+F_{ij}^{(1,0)})=\epsilon_{ijk}[(D_k\phi)^{(1)}+{\cal O}_k^{(1,0)}].
\label{monopolehsu3}
\end{equation}
Now both sides of the equation take values in the enveloping algebra of
$SU(3)$ in the fundamental representation. Hence, eq.~\eqref{monopolehsu3} is
equivalent to the following two equations
\begin{equation}
\begin{array}{l}
a)\quad\Tr\big[(f_{ij}^{(1)}+F_{ij}^{(1,0)})\big]=\epsilon_{ijk}\Tr\big[(D_k\phi)^{(1)}+{\cal
O}_k^{(1,0)}\big],\\
b)\quad\Tr\big[\frac{\lambda^a}{2}(f_{ij}^{(1)}+F_{ij}^{(1,0)})\big]=
\epsilon_{ijk}\Tr\big[\frac{\lambda^a}{2}\big((D_k\phi)^{(1)}+{\cal
O}_k^{(1,0)}\big)\big].\\
\end{array}
\label{idenprojsu3}
\end{equation}
As in the $SU(2)$ case, the equality  $a)$ in the previous
equation only involves the zero order contributions to the field
configurations: $a^{(0)}_i$ and $\phi^{(0)}$ in
eqs.~\eqref{order0} and~\eqref{embed}. Eq.~$a)$ is thus a
constraint on the parameters of the Seiberg-Witten map. Although
$\phi^{(0)}$ has a non-vanishing component along the singlet
generator  $T^s_\beta$, one may show that $a)$ in
eq.~\eqref{idenprojsu3} holds if, and only if, the parameters
$\kappa_2$ and $\lambda_1$ of the Seiberg-Witten map in
eq.~\eqref{SWgen} take the same values as in the $SU(2)$ case
--see eq.~\eqref{solconsist}. Some computations lead to the
conclusion that the equality $b)$ in eq.~ \eqref{idenprojsu3} is
equivalent to the following equation
\begin{equation}
D^{(0)}_i(a'_j)-D^{(0)}_j(a'_i)=\epsilon_{ijk}\,\big(D^{(0)}_k\phi'-i[a'_k,\phi^{(0)}]\big),
\label{su3zeromod}
\end{equation}
where $a'_j$ and $\phi'$ are defined in terms of $a^{(1)}_i$
and $\phi^{(1)}$ by the following identities:
\begin{equation}
\begin{array}{l}
a^{(1)}_i=a'_i-\kappa_1\,\theta^{kl}\,D^{(0)}_i f^{(0)}_{kl}-i\kappa_3\,
{\theta_i}^l\,[D^{(0)}_l\phi^{(0)},\phi^{(0)}]-\kappa_4\,v\,
{\theta_i}^{j}D^{(0)}_j\phi^{(0)}
+\frac{\phi^s}{2\sqrt{3}}{\theta_i}^j\,D^{(0)}_j\phi^{(0)},\\
\phi^{(1)}=\phi'-i\lambda_2\,\theta^{ij}\,[f_{ij}^{(0)},\phi^{(0)}]-
\lambda_3\,v\,\theta^{ij}f_{ij}^{(0)}-\frac{\phi^s}{4\sqrt{3}}\,\theta^{ij}\,f_{ij}^{(0)},
\end{array}
\label{su3bpssol}
\end{equation}
respectively.

Eq.~\eqref{su3zeromod} is defining, modulo gauge
transformations,  the zero modes, or moduli, of the corresponding
ordinary fundamental monopole. Hence, $a'_i$ and $\phi'$ are given
by appropriate linear combinations of the corresponding moduli
with coefficients that depend linearly on $h\theta^{\m\nu}$. This
is completely analogous to what we found in the $SU(2)$ case.
However, we see that now $a^{(1)}_i$ and $\phi^{(1)}$ contain
extra contributions, as compared with those in eq.~\eqref{defone},
coming from the singlet part, $\phi^s\,T_\beta^s$, of
$\phi^{(0)}$. And yet, the complete noncommutative correction to
the ordinary $SU(3)$ BPS fundamental monopoles  is a linear
combination of ordinary zero modes and field redefinitions. Let us
stress that the values of the real coefficients $k_1$, $k_3$, $k_4$,
$\lambda_2$ and $\lambda_3$ that parametrize the ambiguity in the
Seiberg-Witten map corresponding to field redefinitions have no
bearing on the existence of noncommutative BPS solutions. However,
the existence of these noncommutative field configurations demands
$k_2=-\frac{1}{2}$ and $\lambda_1=\frac{1}{4}$, $k_2$ and
$\lambda_1$ parametrizing the ambiguities of the Seiberg-Witten map
that cannot be interpreted neither as field redefinitions nor as
gauge transformations.

In ordinary space-time, there is another natural embedding of
$SU(2)$ into $SU(3)$. This is the embedding along the remaining
positive root $\vec{\beta}_3=\vec{\beta}_1+\vec{\beta}_2$. The embedding of the
ordinary $SU(2)$ monopole in the $SU(2)$ subgroup of $SU(3)$
defined by $\vec{\beta}_3$ has topological charges $(1,1)$ and is not a 
fundamental monopole but rather a two-monopole field configuration
constituted by a fundamental monopole of type $(1,0)$ and another
of type $(0,1)$ superimposed at the same point. The mass and
magnetic charge of this $(1,1)$ two-monopole are the sum of those of its
constituent fundamental monopoles --see~\cite{Weinberg:1979zt} for
further information. Obviously, the noncommutative counterpart of
the previous ordinary two-monopole is given, at first order in
$h\theta^{\mu\nu}$ and if eq.~\eqref{solconsist} holds, by
eq.~\eqref{su3bpssol}, but, now, $\lambda$ is equal to
$v\,\vec{h}\cdot \vec{\beta}_3$ and the generators of $SU(3)$,
$T^a_{\beta_3}$, are defined in terms the eight Gell-Mann
matrices, $\lambda^a$, as follows
\begin{equation}
\text{Triplet:}\,\{T^4,T^5,\frac{1}{2}T^3+\frac{\sqrt{3}}{2}T^8\},
\quad\text{Doublets:}\,\,\{T^1,T^2,T^6,T^7\},\quad\text{Singlet:}\,\,
\frac{\sqrt{3}}{2}T^3-\frac{1}{2}T^8.
\label{basisb3}
\end{equation}
The labels Triplet, Doublets and Singlet refer to the behaviour of
$T^a$, $a=1\dots 8$, under the adjoint action of the $SU(2)$
generators $T^a_{\beta_3}$, $a=1,2,3$.

The noncommutative field configuration we have just constructed
has  topological charges $(1,1)$ and mass  $M_{3}$ equal to
$M_1+M_2$, with $M_1=v\,\vec{h}\cdot\vec{\beta}_1$ and
$M_2=v\,\vec{h}\cdot\vec{\beta}_2$. Further, one may argue that,
as is the case with its ordinary counterpart, there are eight zero
modes, or moduli, associated with it. Indeed, the number of
linearly independent normalizable  zero modes can be obtained by
computing the index of an operator that differs from the
corresponding ordinary operator in ref.~\cite{Weinberg:1979zt} by
terms that are of order one in $h\theta^{\mu\n}$. These terms are to
be considered small continuous perturbations of the ordinary
operator and hence they will not change the value of the index
--this is actually what happens in the case of the chiral anomaly in
ref.~\cite{Martin:2005jy} and for fundamental monopoles. It  is
therefore natural to conclude that the noncommutative BPS
configuration obtained for the root $\vec{\beta}_3$ is made out of two
fundamental noncommutative monopoles: a $\beta_1$-monopole and a
$\beta_2$-monopole.

Finally, it is straightforward to repeat the previous analysis for negatively charged monopoles,
obtained as deformations of the embeddings of the SU(2) anti-monopole along the SU(2) subalgebras
defined by the roots $\vec{\beta}_1,\,\vec{\beta}_2$ and $\vec{\beta}_3$. The same conclusions
are reached as in the case of positively charged monopoles.


\subsection{Noncommutative $SO(5)$ theory and BPS massless monopoles}

In ordinary space-time, when the unbroken gauge group is not the
maximal torus of the broken gauge group, there show up massless
monopoles~\cite{Lee:1996vz}. These objects do not occur as
isolated solutions to the equations of motion,  but manifest
themselves in multi-monopole field configurations as clouds
surrounding massive  monopoles and carrying non-abelian magnetic
charges. The simplest example where these field configurations with
massless monopoles occur is furnished by $SO(5)$  Yang-Mills-Higgs
theory, with $SO(5)$ broken down to $SU(2)\times U(1)$. An
eight-moduli family of BPS solutions was found in
ref.~\cite{Weinberg:1982jh} that contains one fundamental massive
$\beta$-monopole and one massless $\gamma$-monopole. The
Higgs field of this configuration satisfies the boundary condition
$\phi(0,0,z\rightarrow-\infty)=v\vec{h}\cdot \vec{H}$, with
$\vec{h}\cdot\vec{\beta}>0$ and  $\vec{h}\cdot\vec{\gamma}=0$.
$\{\vec{\beta},\quad\vec{\gamma}\}$ is a set of  simple roots of
$SO(5)$. We label the roots of $SO(5)$ as follows:
$\{\pm\vec{\alpha},\pm\vec{\beta},\pm\vec{\gamma},\pm\vec{\mu}\}$,
where
\begin{equation*}
 \vec{\alpha}=(0,1)\quad\vec{\beta}=\Big(-\frac{1}{2},\frac{1}{2}\Big)
 \quad\vec{\gamma}=(1,0)\quad\vec{\mu}=\Big(\frac{1}{2},\frac{1}{2}\Big).
\end{equation*}
To display the BPS two-monopole solution in question some notation is needed.
Let $E_{\pm\delta}$ be the rising and lowering generators of $SO(5)$ defined
by the root $\vec{\delta}$ of the latter. Let ${T^a}_\delta$  denote, for $a=1,2,3$,
the generators of the $SU(2)$ subgroup of $SO(5)$ defined by the root $\vec{\delta}$.
Then, any element, $Q$, of the Lie algebra of $SO(5)$ admits the following decomposition:
\begin{equation*}
 Q=\sum_{a=1}^3{Q(1)^a {T_\alpha}^a}+\sum_{a=1}^3{Q(2)^a {T_\gamma}^a}+
 \tr Q(3) M,\,\,M=\frac{i}{\sqrt{\beta^2}}\left(\begin{array}{cc}
E_\beta & -E_{-\mu}\\
E_\mu & E_{-\beta}
\end{array}\right),
\end{equation*}
where $Q(3)^*=-\sigma_2 Q(3) \sigma_2$, with $\sigma_2$ denoting
the second Pauli matrix. Then, the field configuration constituted
by a massive $\vec{\beta}$-monopole and a massless
$\vec{\gamma}$-monopole has the following components:
$Q(s)^a=a_i(s)^a\,\text{or}\;\phi(s)^a$, $s=1,2$ and $a=1,2,3$,
and $Q(3)=a_i(3)\,\text{or}\,\phi(3)$, with
 \begin{equation}
\begin{array}{c}
\begin{array}{lll}
a_i(1)^a=\epsilon_{aim}\,A(r)\frac{x_m}{r},         &       \phi(1)^a=H(r)\,\frac{x_a}{r},\\
a_i(2)^a=\epsilon_{aim}\,G(r,b)\frac{x_m}{r},       &       \phi(2)^a=G(r,b)\,\frac{x_a}{r},\\
 a_i(3)=\sigma_i\,F(r,b),                            &       \phi(3)=-i I F(r,b),
\end{array}\\
A(r)=\frac{1}{r}-\frac{v}{\sinh(vr)},\quad H(r)=\frac{1}{r}-v\coth(vr),
\quad F(r,b)=\frac{v}{\sqrt{8}\cosh(vr/2)}\,L(r,b)^{1/2},\\
G(r,b)=A(r)L(r,b),\quad
L(r,b)=\big[1+\frac{r}{b}\,\coth(\frac{vr}{2})\big]^{-1},\,\,b>0.
\end{array}
\label{so5ord}
\end{equation}
$\sigma_i$ and $I$ stand for the Pauli matrices and the
$2\times2$ identity matrix, respectively, and $v=2\vec{\beta}\cdot
\vec{h}$. Notice that under the unbroken $SU(2)$ subgroup
furnished by $\vec{\gamma}$, $Q^a(1)$, $Q^a(2)$ and $Q(3)$
transform as three singlets, a triplet and a complex doublet. The
$SU(2)$ triplet $Q^a(2)$ decays as $1/r$ in the region
$1/v\lesssim r \lesssim b$. This is the non-abelian cloud
representing classically the massless monopole which is charged
under the unbroken $SU(2)$ --for further discussion, see
refs.~\cite{Lee:1996vz, Weinberg:2006rq}.

The purpose of this subsection is to see whether, at first order
in $h\theta^{\m\n}$, there exist
solutions to the noncommutative BPS equation, $B_i-D_i\Phi=0$, that
are formal power series in $h\theta^{\mu\nu}$ and that go to the
field configuration in eq.~\eqref{so5ord} as
$h\theta^{\mu\nu}\rightarrow 0$. We shall assume that the
generators of $SO(5)$ are in the fundamental representation. The
contribution, at first order in $h\theta^{\m\n}$, to the non-abelian
BPS equation reads
\begin{equation*}
E\equiv f_{ij}^{(1)}+F_{ij}^{(1,0)}-\epsilon_{ijk}[(D_k\phi)^{(1)}+{\cal O}_k^{(1,0)}]=0.
\end{equation*}
The notation is the same as in subsection 2.1, but now $E$ belongs
to the enveloping algebra of $SO(5)$ in the fundamental
representation. In the previous cases, since we were dealing with
$SU(N)$ groups in the fundamental representation, any element of
the  enveloping algebra  could be expressed as a linear
combination of the generators of the Lie algebra and the identity;
this is no longer the case now. The generators of the Lie algebra
of SO(5) in the fundamental representation can be taken as pure
imaginary hermitian---and therefore antisymmetric---matrices; then,
the enveloping algebra includes
also all the real symmetric matrices. It is possible to construct
a basis $\{R^a\}$ of the enveloping algebra of $SO(5)$ in the
fundamental representation that is made out of the generators of
$SO(5)$, $\{T^a\},\,a=1\dots10$ and a basis $\{S^a\},\,a=1\dots
15$ of the real symmetric matrices. The whole basis can be made
orthogonal with respect to the trace operation: $\Tr\, R^a
R^b\propto\delta^{ab}$. Using this orthogonal basis, the previous
equation can be projected out onto a given element of the former
just by first multiplying the latter by the element in question
and, then, taking traces:
\begin{equation*}
E=0\quad\Leftrightarrow\quad\,\Tr\, S^a \,E=0,\;\forall a=1\dots
15\quad\text{and}\quad \Tr\, T^a\,E=0,\;\forall a=1\dots 10.
\end{equation*}
Since the trace of an antisymmetric matrix times a symmetric one
vanishes, it turns out that $f^{(1)}_{ij}$ and $(D_k\phi)^{(1)}$
drop out from $\Tr\, S^a \,E=0$. Hence, only the ordinary field
configuration enters the equations $\Tr\, S^a \,E=0$, which are
thus turned into the following constraint on the parameters of
the Seiberg-Witten map:
\begin{equation}
\begin{array}{l}
\Tr \,S^a[F_{ij}^{(1,0)\,\text{st}}-\epsilon_{ijk}{\cal
O}_k^{(1,0)\,\text{st}}]=-
\Tr S^a[D^{(0)}_i(\,\kappa_2\,{\theta_j}^k\{(D_k\phi)^{(0)},\phi^{(0)}\})-(i\leftrightarrow j)]\\
\phantom{\Tr[F_{ij}^{(1,0)\text{st}}\mp\epsilon_{ijk}{\cal
O}_k^{(1,0)\text{st}}]=}
+\epsilon_{ijm}\Tr\,S^a\,D_m^{(0)}\,[\,\lambda_1\,\theta^{kl}\{f^{(0)}_{kl},\phi^{(0)}\}].
\end{array}
\label{sameconst}
\end{equation}
The reader is referred to subsection 3.2 for notation. The
superscript "st" shows that the corresponding object is computed
by using the standard form  --all free parameters set to zero-- of
the Seiberg-Witten map. Now, substituting eq.~\eqref{so5ord} in
eq.~\eqref{sameconst}, one ends up with the conclusion that the
resulting equation holds if, and only if, $\kappa_2$ and
$\lambda_1$ take the values quoted in eq.~\eqref{solconsist}.

It remains to solve $\Tr\,T^a\,E=0$. Since $\{T^a,T^b\}$ is a
symmetric matrix, the previous equation boils down to
\begin{equation}
\begin{array}{l}
D^{(0)}_i(a'_j)-D^{(0)}_j(a'_i)=\pm\epsilon_{ijk}\,\big(D^{(0)}_k\phi'-i[a'_k,\phi^{(0)}]\big)\\
a'_i=a^{(1)}_i+\kappa_1\,\theta^{kl}\,D^{(0)}_i
f^{(0)}_{kl}+i\kappa_3\,{\theta_i}^l\,
[(D_l\phi)^{(0)},\phi^{(0)}]+\kappa_4\,v{\theta_i}^j D^{(0)}_j\phi^{(0)},\\
\phi'=\phi^{(1)}+i\lambda_2\,\theta^{kl}\,[f_{kl}^{(0)},\phi^{(0)}]+\lambda_3\,v\,\theta^{ij}f_{ij}^{(0)},
\end{array}
\label{correct}
\end{equation}
where  $(a_i^{(0)},\phi^{(0)})$ denotes the ordinary field
configuration of eq.~\eqref{so5ord}. Hence, the
first-order-in-$h\theta^{\mu\nu}$ BPS corrections to the ordinary
field configuration are given, again,  by the terms of the Seiberg-Witten
map associated to field redefinitions plus $\theta$-dependent linear
combinations of the ordinary zero modes.

One may care to compute the first-order-in-$h\theta^{\mu\nu}$ BPS
corrections to ordinary fundamental monopoles for $SO(5)$ in the
fundamental representation. Proceeding as in the previous
paragraphs one concludes that they exist if, and only if,
eq.~\eqref{solconsist} holds, and that they are given by
eq.~\eqref{correct}, if $(a_i^{(0)},\phi^{(0)})$  denotes
now the ordinary fundamental monopoles. Let us stress that we have shown
that, for $SU(2)$ and $SO(5)$ in their fundamental reperesentations,
$a^{(1)},\phi^{(1)}$ are given by the same type of corrections. A result that
has its origin partially in  the fact that for both groups
$\Tr\, T^a\{T^b,T^c\}=0$.  Notice that $\Tr\, T^a\{T^b,T^c\}\neq 0$ for $SU(3)$.


\section{Static solutions to the BPS Yang-Mills-Higgs equations at first order in $h\theta^{\mu\nu}$}

In the previous section, we have seen that for some gauge simple
groups only if the parameters labeling the ambiguity of the
Seiberg-Witten map are appropriately chosen there exist
noncommutative BPS (multi-)monopoles that are power series in
$h\theta^{\m\n}$ and that go to a given ordinary BPS
(multi-)monopole configuration as $h\theta^{\m\n}\rightarrow 0$.
The next question to ask is whether given an ordinary BPS
(multi-)monopole configuration there exists for any value of
$\kappa_2$ and $\lambda_1$ a solution to the noncommutative
Yang-Mills-Higgs equations in the BPS limit that has the following
properties: it is static, it is a power series in $h\theta^{\m\n}$
and it goes to the given ordinary BPS (multi-)monopole
configuration as $h\theta^{\m\n}\rightarrow 0$. Notice that the
noncommutative BPS equations had contributions that were
proportional to the identity matrix, and this was part of the
problem, whereas in the noncommutative Yang-Mills-Higgs equations in
the BPS limit for simple groups, which are displayed in
eq.~\eqref{NCYMH}, no contribution of that sort occurs.

    \subsection{SU(2) case}

    At zero order in $h\theta^{\m\n}$, the equations of motion are the ordinary ones and hence they are
satisfied by $a^{(0)}_i,\phi^{(0)}$ --we use the notation of
eq.~\eqref{expfields}-- given by ordinary BPS (multi-)monopole
configurations.
 Let us choose the gauge $a_0=0$. After carrying out some simplifications, it can be shown that
 the contributions to eqs.~\eqref{NCYMH}, at first order in $h\theta^{\m\n}$ and for time
 independent field configurations, read
\begin{equation}
\begin{array}{l}
    D_i D_i\phi'-i D_i[a'_i,\phi]-i[a'_i,D_i\phi]=0,\\
    D_i (D_i a'_j-D_j a'_i)-i[a'_i,f_{ij}]+i[\phi',D_j\phi]+i[\phi,D_j\phi'-i[a'_j,\phi]]=0,\\
a'_j=a^{(1)}_j+\kappa_1\,\theta^{kl}\,D_j f_{kl}+i\kappa_3\,{\theta_j}^l\,[(D_l\phi),\phi]+v\kappa_4\,{\theta_j}^lD_{l}\phi,\\
\phi'=\phi^{(1)}+i\lambda_2\,\theta^{kl}\,[f_{kl},\phi]+ v\,\lambda_3\,\theta^{ij}f_{ij}\\
\label{su2eom}
\end{array}
\end{equation}
where $D_i=D_i^{(0)}=\partial_i-i[a^{(0)}_i,\phantom{a^{(0)}_i}]$,
$f_{ij}=f^{(0)}_{ij}$,
$a_i=a^{(0)}_i$ and $\phi=\phi^{(0)}$, $a^{(0)}_i$ and $\phi^{(0)}$ being the
fields defining the ordinary BPS $SU(2)$ (anti-)monopole. It is natural to  look for $a'_i$ and $\phi'$ such that
\begin{equation}
a'_i(\vec{x})\sim \frac{1}{|\vec{x}|^2}\quad\text{and}\quad \phi'(\vec{x})\sim\frac{1}{|\vec{x}|^2}
\quad\text{as}\quad|\vec{x}|\rightarrow\infty.
\label{primeboundary}
\end{equation}
Note that one readily deduces from eq.~\eqref{ordboundary} that the terms that go
with $\kappa_1$, $\kappa_2$ and $\kappa_3$ and $\lambda_2$ and $\lambda_3$
in eq.~\eqref{su2eom} satisfy the previous boundary conditions and that these boundary
conditions guarantee that there will be  no $\theta^{\mu\n}$
dependent contributions to the magnetic charge defined in eq.~\eqref{ncmagcharge}.
The latter contributions would put into jeopardy the interpretation of
the magnetic charge as a topological quantity.

Let us analyse the equations for $\phi'$ and $a_i'$. Using the
fact that $a^{(0)}$ and $\phi^{(0)}$ satisfy the ordinary BPS
equations, the first equation in eq.~(\ref{su2eom}) leads to
\begin{equation}
D_i(D_i\phi'-i[a_i',\phi]\mp\epsilon_{ijk}D_j a'_k)=0.
\label{zeromphi}
\end{equation}
    Introducing the four-vector fields in three dimensions $\bar{a'}_\mu=(a'_i,\phi')$ and
$\bar{a}_\mu=(a^{(0)}_i,\phi^{(0)})$, one may cast eq.~(\ref{zeromphi})
into the form
\begin{equation*}
\bar{D}_\mu (\bar{D}_\mu \bar{a'}_4-\bar{D}_4 \bar{a'}_\mu\mp\epsilon_{\m 4 \r\s}\bar{D}_\r \bar{a'}_\s)=0.
\end{equation*}
    This equation is of the type $\bar{D}_\mu \bar{X}_{\m4}=0$ with $\bar{D}$ in the background
of a self-dual field $\bar{a}_\mu$ and with $\bar{X}_{\mu 4}$
being self-dual. Using the techniques
 in ref.~\cite{Martin:2005vr}, one may show that the only normalizable solutions
 to this equation are those satisfying $X_{i4}=0$. Notice that $\bar{D}_\mu \bar{a'}_4-\bar{D}_4
 \bar{a'}_\mu\mp\epsilon_{\m 4 \r\s}\bar{D}_\r \bar{a'}_\s$ must be a smooth function
 of $\vec{x}$ such that it decays as $1/|\vec{x}|^2$ as
$|\vec{x}|\rightarrow\infty$ and, hence, normalizable in three
dimensions. Now,  $\bar{X}_{\mu 4}=0$ yields
\begin{equation}
D_i\phi'-i[ a_i',\phi]\mp\epsilon_{ijk}D_j a'_k=0.
\label{zeromphi2}
\end{equation}
This is precisely the equation of the zero modes in the background
of an ordinary BPS $SU(2)$ (anti-)monopole. Going back to the second
equation in eq.~(\ref{su2eom}),  using the result in
eq.~(\ref{zeromphi2}) and the condition
$f_{ij}=\pm\epsilon_{ijk}D_k\phi$,  we arrive at
\begin{equation*}
D_i(D_i a'_j-D_j a'_i\mp i[\phi,\epsilon_{ijl} a'_l]\mp\epsilon_{ijl}D_l\phi')=0,
\end{equation*}
which is automatically satisfied if eq.~(\ref{zeromphi2}) holds.
We therefore conclude that $\phi',\,a'_i$ which satisfy the
boundary conditions in eq.~\eqref{primeboundary} --see comments
below eq.~\eqref{primeboundary}-- are just linear combinations of
the zero modes of the ordinary BPS (anti-)monopole with
$\theta^{\m\n}-$dependent coefficients. We thus conclude that
there are solutions, for $SU(2)$ and at first order in
$h\theta^{\mu\nu}$, to the noncommutative  Yang-Mills-Higgs
equations in the BPS limit, whatever the value of the parameters
labeling the ambiguity of the Seiberg-Witten map. These solutions
are given by the field redefinitions of the ordinary BPS (anti-)monopole
furnished by the Seiberg-Witten map plus some appropriate linear
combinations of the ordinary $SU(2)$ zero
 modes.


\subsection{SU(3) case}

Let ($a^{(0)}_i,\;\phi^{(0)}$) denote the ordinary BPS monopole
and two-monopole solutions in eq.~\eqref{order0}. Let
$D_i=\partial_i-i[a^{(0)}_i,\phantom{a^{(0)}_i}]$,
$f_{ij}=\partial_i a^{(0)}_j-\partial_j a^{(0)}_i-i[a^{(0)}_i,
a^{(0)}_j]$, $\phi=\phi^{(0)}$, and let $a'_j$ and $\phi'$ be
given by
\begin{equation}
\begin{array}{l}
a'_k=a^{(1)}_k+\kappa_1\,\theta^{ij}\,D_k f_{ij}+
\frac{\kappa_2\phi^s}{\sqrt{3}}\,{\theta_k}^j D_j\phi+
\frac{\kappa_2}{\sqrt{3}}{\theta_k}^j\,(D_j\phi)^{a}\phi^{ta} T^s_\beta
+i\kappa_3\,{\theta_k}^l\,[(D_l\phi),\phi]\\
\phantom{a'_k=a^{(1)}_k}+\kappa_4\,v\,{\theta_k}^j D_j\phi\\
\phi'=\phi^{(1)}+\frac{\lambda_1\phi^s}{\sqrt{3}}\theta^{ij}\,f_{ij}+
\frac{\lambda_1}{\sqrt{3}}\,\theta^{ij}f_{ij}^{a}\,\phi^{ta} T^s_\beta+i\lambda_2\,\theta^{ij}\,
[f_{ij},\phi]+\lambda_3\,v\,\theta^{ij}\,f_{ij}.\\
\end{array}
\label{su3eom}
\end{equation}
See subsection 3.2 for notation.
 Then, for $SU(3)$, the first order in $h\theta^{\m\n}$ contribution to the
noncommutative Yang-Mills-Higgs equations of eq.~\eqref{NCYMH} in
the gauge $a_0=0$ and for time independent field configurations
reads
\begin{equation}
\begin{array}{l}
\Tr\, T^a\big[2D_j D_j\phi'-2iD_j[a'_j,\phi]-2i[a'_j,D_j\phi]\big]=\\
-\Tr\,T^a \theta^{ij}\Big[-\frac{1}{2}D_m\{D_m\phi,f_{ij}\}-
D_j\{D_m\phi,f_{mi}\}-D_m\{D_j\phi,f_{mi}\}\Big],\\[10 pt]
\Tr\, T^a\big[
-2D_i(D_i\,a'_k-D_k\, a'_i)+2i[ a'_i,f_{ik}]-2i[\phi',D_k\phi]-2i[\phi,-i[ a'_k,\phi]+D_k\,\phi']\big]=\\
-\Tr\, T^a\Big[{\theta^i}_k\big(-\frac{1}{4}D_i\{f_{mn},f_{mn}\}-
D_m\{f_{n i},f_{m n}\}-\frac{1}{2}D_i\{D_m\phi,D_m\phi\}+D_m\{D_i\phi,D_m\phi\}\big)\\
 +\theta^{ij}\big(\frac{1}{2}D_m\{f_{mk},f_{ij}\}+D_i\{f_{m j},f_{k m}\}-
 D_m\{{f}_{mi},f_{k j}\}-D_i\{D_j\phi,D_k\phi\}\big)\Big].\\
\end{array}
\label{tracepro}
\end{equation}

The non-zero traces that occur on the r.h.s of both equalities in
eq.\eqref{tracepro} are of the type
$\Tr\,T^a\{T_\beta^b,T_\beta^c\}$. Since $\{T_\beta^b,T_\beta^c\}$
behaves as a singlet under the $SU(2)$ Lie algebra generated by
$\{T_\beta^c\}_{c=1,2,3}$,  the r.h.s. is only nonzero if $T^a$
is the SU(2) singlet generator. The corresponding l.h.s of the
equations will pick up only the components of $a'_i,\,\phi'$ along
the singlet, since for the basis in eqs.~\eqref{bases} and
eq.~\eqref{basisb3}, $\Tr\,T^a T^b=\frac{1}{2}\delta^{ab}$ holds
and the SU(2) subalgebra defined by the root $\vec{\beta}$ acts
irreducibly on the specified representations. Hence, the equations
for the components of $\phi'$ and $a_k'$ along the singlet
decouple from the rest. We can  express $\phi'$ and $a'_i$ as
follows: $\phi'=\phi'^s+\phi''$, with $\phi'^s$ being the
component along the SU(2) singlet, and analogously
$a'_i=a_i'^s+a_i''$.

        Let us first analyse the equations for $\phi''$ and $a_i''$.
        In this case the r.h.s. of
the equalities in eq.~\eqref{tracepro} vanishes, so that  we are
left with the same equations given by the first two lines of
eq.~(\ref{su2eom}), whose solutions for the boundary conditions of
eq.~\eqref{primeboundary} are  given by the zero modes in the
background of the ordinary $SU(3)$ BPS (two-)monopole.\par
    It remains to solve the equations for the components, $\phi'^s,\,a_i'^s$,
    along the singlet. For each SU(2) embedding in section 3.2, we choose the corresponding basis
in eq.~(\ref{bases}) and eq.~\eqref{basisb3} and take $T^a$ in
eq.~(\ref{su3eom}) to be the corresponding singlet generator. Now, for
each basis the property $\Tr\,
T_\beta^{s}\{T_\beta^a,T_\beta^b\}=\frac{1}{2\sqrt{3}}\delta^{ab}$
holds, so that we end up with the following equations:
\begin{equation*}
\begin{array}{l}
\partial_i\partial_i\phi'^s=\frac{1}{2\sqrt{3}}\,\theta^{ij}\,\Big[\frac{1}{2}
\partial_k[(D_k\phi)^a(f_{ij})^a]+\partial_j[(D_k\phi)^a(f_{ki})^a]+\partial_k[(D_j\phi)^a(f_{ki})^a]\Big],\\
\partial_i \partial_i a'^s_j-\partial_j\partial_i a'^s_i=
\frac{1}{2\sqrt{3}}{\theta^i}_j\Big[-\frac{1}{4}\,\partial_i[f^a_{mn}f^a_{mn}]-
\partial_m[f^a_{n i}f^a_{mn}]-\frac{1}{2}\,\partial_i[(D_m\phi)^a (D_m\phi)^a]\\
 +\partial_m[(D_i\phi)^a(D_m\phi)^a]\Big]\!+\!\frac{1}{2\sqrt{3}}\,\theta^{ik}
 \Big[\frac{1}{2}\,\partial_m[f^a_{mj}f^a_{ik}]\Big]
+\partial_i[f^a_{mk}f^a_{jm}]-\partial_m[f^a_{mi}f^a_{jk}]-\partial_i[(D_k\phi)^a (D_j\phi)^a]\Big].
\end{array}
\end{equation*}
The computation of the r.h.s of both equations for the field
configurations in eqs.~(\ref{order0}) yields
\begin{equation*}
\begin{array}{l}
\partial_i\partial_i\phi'^s=\theta^{ij}\epsilon_{ijk}\, x^k f(r),\\
\partial_i \partial_i a'^s_j-\partial_j\partial_i a'^s_i=2{\theta^i}_j x^i f(r),\\
f(r)=\frac{1}{2\sqrt{3}}\big[\frac{1}{2r}\frac{d}{dr}H'^2+\frac{1}{r}\frac{d}{dr}\big(\frac{K'}{r}\big)^2\big].
\end{array}
\end{equation*}
    The general solution to each of these equations is the sum
    of a particular solution plus a solution to the
homogeneous equation. The homogeneous equation for $\phi'^s$ has no
smooth solution that vanishes at infinity, while the homogeneous
equation for $a_i'^s$ has as non-singular solutions total derivatives which
are equivalent to gauge transformations. Therefore we just need to find non-singular
particular solutions that respect the boundary conditions. Choosing the following ans\"atze,
\begin{equation}
    \phi'^s=\theta^{ij}\,\epsilon_{ijk}x^k g(r),\quad a'^s_j={\theta^i}_j x^ih(r),
\label{sol1}
\end{equation}
one finds  the following solution
\begin{equation}
\begin{array}{l}
g(r)=\frac{1}{2}\,h(r)=\frac{1}{4\sqrt{3}}\,\frac{H(1-K)(3-K)}{r^3}=
-\frac{1}{16\sqrt{3}r^4}\,\text{csch}^3(r\lambda)[r\lambda\text{cosh}(r\lambda)(1+4(r\lambda)^2)\\
\phantom{g(r)=-\frac{1}{2}\,f(r)=}-r\lambda\text{cosh}(3r\lambda)
+2\text{sinh}(r\lambda)(-1-2(r\lambda)^2+\text{cosh}(2r\lambda))],
\end{array}
\label{sol2}
\end{equation}
where $\lambda=v\,\vec{h}\cdot\vec{\beta}$.
    Putting it all together and  realizing that the singlet contributions
    to $a'_i$ and $\phi'$ in eq.~(\ref{su2eom}) are proportional to the previously given
    $g(r)$, one ends up with the following  family of static solutions to the
     first order in $h\theta^{\mu\nu}$ equations of motion:
\begin{equation}
\begin{array}{l}
\phi^{(1)}=\delta\phi^{(0)}+(1-4 \lambda_1)\theta^{ij}\epsilon_{ijk}x^k g(r)
T_\beta^s-\frac{\lambda_1\,\phi^s}{\sqrt{3}}\theta^{ij}\,f_{ij}-i \lambda_2\,
\theta^{ij}\,[f_{ij},\phi]-\lambda_3\,v\,\theta^{ij}\,f_{ij}\\
a_i^{(1)}\!=\!\delta a_i^{(0)}\!+\!(4\kappa_2\!+\!2){\theta^j}_i x^j g(r)
T^s_\beta-\kappa_1\,\theta^{kl}\,D_i f_{kl}-\frac{\kappa_2\phi^s}
{\sqrt{3}}\,{\theta_i}^j D_j\phi-i\kappa_3{\theta_i}^j[D_j\phi,\phi]\\
\phantom{a_i^{(1)}\!=}-\kappa_4\,v\,{\theta_i}^j\,D_j\phi.\\
\end{array}
\label{su3nbps}
\end{equation}
$\delta\phi^{(0)}$ and $\delta a_i^{(0)}$ denote any linear
combination of the zero modes of the corresponding ordinary BPS
configuration with coefficients that depend linearly on
$h\theta^{\m\n}$. $D_i$, $f_{ij}$ and $\phi$ have been defined at
the beginning of this subsection. The solutions reported in
eq.~\eqref{su3nbps} are well behaved at $r=0$ and the behaviour at
infinity is such that the noncommutative corrections respect the
ordinary boundary conditions and do not contribute to the magnetic
charge. When $\kappa_2=-\frac{1}{2},\,\lambda_1=\frac{1}{4}$,
values for which there exist solutions to the noncommutative BPS
equations, the singlet contributions vanish and we recover the
field configurations that solve the noncommutative BPS equations
for $SU(3)$ --see eq.~\eqref{su3bpssol}.

The solution in eq.~\eqref{su3nbps}, which exists for any value of
the parameters of the Seiberg-Witten map defined in
eq.~\eqref{SWgen} --with $w=0$, of course--, constitutes a
noncommutative deformation of the ordinary  BPS field
configuration obtained by embedding the ordinary BPS $SU(2)$
monopole along the root $\vec{\beta}$, with $\vec{\beta}=
\vec{\beta}_i$, $i=1,2,3$. The mass, $M_{\beta}$, of the complete
static field configuration, which in general is a noncommutative
non-BPS field configuration, is obtained by substituting
$a_i=a^{(0)}+h a_i^{(1)}$ and $\phi=\phi^{(0)}+h \phi^{(1)}$ in
eq.~\eqref{Hambogo}. After a little algebra one ends up with a
number of integrals that have to be worked out numerically. The
final answer for $M_{\beta}$ is then given by
\begin{equation}
M_{\beta}=M_{\text{ordinary}}+0.10274\,\, h^2\theta^{ij}\theta^{ij}\,
\lambda^5\, \Big[\Big(\kappa_2+\frac{1}{2}\Big)^2+2\Big(\lambda_1-\frac{1}{4}\Big)^2\Big]+
O(h^3\theta^3).
\label{nonBPSmass}
\end{equation}
$M_{\text{ordinary}}=4\pi\lambda$ is the ordinary mass and
$\lambda=v\,\vec{\beta}\cdot \vec{h}$, with $\vec{h}$ defined in
eq.~\eqref{hsu3} and such that $\vec{\beta}_i\cdot \vec{h}>0$,
$\forall i$. Recall that $h^2$ does not denote
$\vec{h}\cdot\vec{h}$. Notice that the quadratic contributions in
$h\theta^{\mu\nu}$ to $M_{\beta}$ are not affected by the
quadratic contributions in   $h\theta^{\mu\nu}$ to field
configurations since $(a^{(0)}_i,\phi^{(0)})$ satisfies the
ordinary BPS equations.

Let $\vec{\beta}_1=(1,0)$ and
$\vec{\beta}_{2}=(-\frac{1}{2},\frac{\sqrt{3}}{2})$; then, $\vec{h}$ is
given by $\vec{h}=(\omega,\sqrt{1-\omega^2}),
\,0<\omega<\frac{\sqrt{3}}{2}$, for  $\vec{\beta}_i\cdot
\vec{h}>0$, $i=1,2$.

Now, if $\vec{\beta}=\vec{\beta}_1$, then, eq.~\eqref{su3nbps}
corresponds generically to a noncommutative non-BPS monopole with
topological vector charge $(1,0)$. If $\vec{\beta}=\vec{\beta}_2$,
then eq.~\eqref{su3nbps} corresponds generically to a
noncommutative non-BPS monopole with topological vector charge
$(0,1)$. Finally, if
$\vec{\beta}=\vec{\beta}_3=\vec{\beta}_1+\vec{\beta}_2$ we have
generically a noncommutative non-BPS two-monopole configuration
with topological charges $(1,1)$. We do not think  that --unless
$\kappa_2=-\frac{1}{2},\,\lambda_1=\frac{1}{4}$ holds-- the
two-monopole configuration is stable. Indeed a little algebra
reveals that $M_{\beta_3}>M_{\beta_1}+M_{\beta_2}$, if
$\kappa_2=-\frac{1}{2},\,\lambda_1=\frac{1}{4}$ is not satisfied
and $0<\omega<\frac{\sqrt{3}}{2}$. Indeed,
\begin{equation*}
 M_{\beta_3}-(M_{\beta_1}+M_{\beta_2})=0.10274\, h^2\theta^{ij}\theta^{ij}v^5\,
 \Big[\Big(\kappa_2+\frac{1}{2}\Big)^2+2\Big(\lambda_1-\frac{1}{4}\Big)^2\Big]
\Big[\frac{15}{16}\,\omega\,(3-4\,\omega^2)\Big]>0.
\end{equation*}
This inequality suggests that the noncommutative character of space
gives rise to a repulsive interaction between the
$\vec{\beta}_1$-monopole and the $\vec{\beta}_2$-monopole that constitute
the field configuration for $\vec{\beta}_3$. Hence, an infinitesimal
disturbance of the static configuration with mass $M_{\beta_3}$
will make the object decay into a system constituted by two
infinitely separated noncommutative non-BPS monopoles, one of type
$\vec{\beta}_1$ and the other of type $\vec{\beta}_2$. Notice that the latter
two-monopole system has mass equal to
 $M_{\beta_1}+M_{\beta_2}$ and belongs to the topological class of the
non-commutative non-BPS $\vec{\beta}_3$-field configuration. This result
casts doubts on the stability of other  non-BPS multi-monopole
configurations for simple gauge groups.

	The extension to the case of negatively charged monopoles is once again trivial; anti-monopole
configurations pick up a minus sign in the term proportional to $(1-4 \lambda_1)$ in eq.~\eqref{su3nbps} and their energy
is equal to that of their positively charged partners.

\subsection{SO(5) case}

    As in the SU(2) case --but not for $SU(3)$--, the traces of the type
$\Tr\,T^a\{T^b,T^c\}$ vanish. We are thus left precisely with the
equations that one finds in eq.~\eqref{su2eom}. Repeating the
analysis made below eq.~\eqref{su2eom}, we arrive at the same
result, i.e., that the first order in $h\theta^{\m\n}$
deformations of the ordinary field configurations are given by the
field redefinitions determined by the Seiberg-Witten map plus
solutions to the zero mode equations in the background of the
ordinary monopole.

\section{Summary, conclusions and outlook}

For three specific gauge groups --$SU(2)$, $SU(3)$ and $SO(5)$--
in their fundamental representations, we have discussed  the
existence of monopole and some two-monopole field configurations
in noncommutative Yang-Mills-Higgs theories in the BPS limit. We
have looked for field configurations that are formal power series
in $h\theta^{\m\n}$ and worked at first order in $h\theta^{\m\n}$.
We have considered a commutative time and the most general
Seiberg-Witten map that leads to an action that, in the gauge
$a_0=0$, contains only first order time derivatives of the fields
and is a quadratic functional of them. We have shown that there is
no monopole solution to these BPS equations unless  two a priori
free parameters of the Seiberg-Witten map are tuned to two
concrete values --see eq.~\eqref{SWgen} and~\eqref{solconsist}.
These free parameters are those free parameters that are not
related with field redefinitions nor with gauge transformations.
The same state of affairs was met when studying the two-monopole
solution that in the limit $h\theta^{\m\n}\rightarrow 0$ goes to
the ordinary $\beta_3$-two-monopole solution of $SU(3)$ and the
noncommutative field configuration that in that very limit yields
the ordinary one-massive-one-massless two-monopole solution of
$SO(5)$. We then showed that whatever the values of the parameters
of our Seiberg-Witten map the noncommutative Yang-Mills-Higgs
equations admit, at first order in $h\theta^{\m\n}$,  monopole
field configurations that solve them and have the same magnetic
charge --although for $SU(3)$ they have different mass-- as the
ordinary  monopoles they go to in the limit
$h\theta^{\m\n}\rightarrow 0$. For $SU(2)$ and $SO(5)$ the first
order in $h\theta^{\m\n}$ corrections correspond to field
redefinitions of the corresponding ordinary object. This is not so
for $SU(3)$. In this case the masses of the field configurations have
contributions that depend quadratically on $\theta^{\m\n}$, so
that the mass of the static
$\vec{\beta}_3=\vec{\beta}_1+\vec{\beta}_2$ field configuration is
larger than the sum of the masses of its constituents: the
$\vec{\beta}_1$-monopole and the $\vec{\beta}_2$-monopole,
$\vec{\beta}_1$ and $\vec{\beta}_2$ being given simple roots of
$SU(3)$. This static $\vec{\beta}_3$ non-BPS field configuration
seems to be unstable.

Let us now state the main conclusions of this paper.  First, at
first order in $h\theta^{\m\n}$, there are BPS monopole solutions
in noncommutative $SU(2)$, $SU(3)$ and $SO(5)$ Yang-Mills-Higgs theory
provided the Seiberg-Witten map is appropriately chosen. This is in sharp
contrast with the instanton case, where no solutions to the noncommutative
self-duality equations could be found already at first order in
$h\theta^{\m\n}$ --see ref.~\cite{Martin:2005vr}. Second,
the parameters $\kappa_2$ and $\lambda_1$ of the Seiberg-Witten
map in eq.~\eqref{SWgen} have physics in them. Indeed, the
properties of the moduli space of the Yang-Mills-Higgs equations
depend on their values: if they take the values of
eq.~\eqref{solconsist},  the elements of the moduli space are BPS objects,
and if they do not, they are non-BPS elements. Notice that the massess of
generally non-BPS $SU(3)$ monopoles depend on  $\kappa_2$ and
$\lambda_1$, see eq.~\eqref{nonBPSmass}. For simple gauge groups,
the fact that the value of the parameters labeling the ambiguity in the
Seiberg-Witten map which is not related to field redefinitions nor
to gauge transformations may have physical consequences is an
issue which cannot be overlooked when considering the
phenomenological applications of the noncommutative theories
constructed within the formalism of refs.~\cite{Madore:2000en,
Jurco:2001rq}. Third, for generic values of $\kappa_2$ and
$\lambda_1$ noncommutative multi-monopole  solutions may become
unstable even if they deform ordinary BPS multi-monopole
configurations.

There are many directions in which the piece of research presented
in this paper can be further developed. We shall mention just a
few of them. First, the computation of the corrections at second
order in $h\theta^{\m\n}$ to the (multi-)monopole field
configurations worked out here. We show in the appendix that, at
variance with the case of instantons --see
ref.~\cite{Martin:2005vr}, Derrick's theorem poses no obstruction
on the existence --for $a_0=0$-- of static field configurations that
solve the equations of motion at second order in $h\theta^{\m\n}$.
Second, it will be interesting to consider  other representations and other
gauge groups. Notice that the field equations take values in the
enveloping algebra of the gauge group, so choosing a
representation may have physical consequences. Third, it is very much needed to
 analyse the question of the stability of non-BPS multi-monopole 
configurations, such as the configuration of eq.~\eqref{su3nbps}, by using the methods of ref.~\cite{ Brandt:1979kk}. Finally, it is a
pressing need to construct supersymmetric generalizations of the
noncommutative models presented here. In ordinary space-time, BPS
monopoles unavoidably occur in some of these theories, so one
wonders whether extended supersymmetry has any bearing on the value of
the parameters of the Seiberg-Witten map and, in particular, if
the values for $\kappa_2$ and $\lambda_1$ in
eq.~\eqref{solconsist} are dictated by supersymmetry.

\section{Acknowledgments}

This work has been financially suported in part by MEC through grant
FIS2005-02309. The work of C. Tamarit has also received financial support
from MEC through FPU grant AP2003-4034. C. Tamarit should like to thank
Dr. Chong-Sun Chu for valuable conversations and the Department of
Mathematical Sciences of the University of Durham, United Kingdom, where part
of this work was carried out, for its kind hospitality.

\section{Appendix: Solutions at higher order in $\theta$ and Derrick's theorem}

In ref.~\cite{Martin:2005vr}, after obtaining the most general
solution to the  noncommutative equations of motion for the
first-order-in-$\theta^{\m\n}$ deformations of the BPST instanton
in noncommutative SU(3) Yang-Mills theory, it was shown by
studying the behaviour of the action under dilatations up to order
$h^2\theta^2$ --i.e., by using Derrick's
theorem~\cite{Derrick:1964ww}-- that there were no solutions that
rendered the action stationary at this order. This conclusion
could be reached because the order  $h^2\theta^2$ constraints on
the action evaluated at the solution to the equations of motion
depended only on the contributions to the field configuration that
were of order $h^0\theta^0$ and $h^{1}\theta^{1}$. In the case
studied here, this does not happen chiefly due to the fact that we
are extremizing the Hamiltonian, which is dimensionful, rather
than dimensionless as the action is, and the Higgs and gauge field
have different scaling behaviours.

    As suits our purposes, we shall choose the gauge $a_0=0$. Proceeding  as in ref.~\cite{Martin:2005vr},
    we shall study the behaviour of the Hamiltonian under infinitesimal dilatations of
    any of  the (multi-)monopole solutions, $(a_i(\vec{x}), \phi(\vec{x}))$, to
the noncommutative Yang-Mills-Higgs equations found in this paper,
those infinitesimal dilatations preserving the boundary conditions
satisfied at infinity by the (multi-)monopole solution:
\begin{equation}
    a'_i=\lambda a_i(\lambda \vec{x}),\quad \phi'=\phi(\lambda\vec{x}),\quad\lambda=1+\delta\lambda.
\label{dilat}
\end{equation}
    The Hamiltonian for an arbitrary Seiberg-Witten map that yields an action of first
    order in time derivatives is given by eq.~(\ref{Hcan}). We
want to obtain the scaling properties of the different
contributions to ${\cal H}$, taking into account eq.~\eqref{dilat}
and dimensional considerations. Because $a_\mu$ and $\phi$ scale
in a different way, contributions to the Hamiltonian expanded in
terms of ordinary fields at a given order in $h\theta^{\m\n}$ will
scale differently depending on the number of $\phi$ fields they
have. In the case of the Standard Seiberg-Witten map --for its
definition, see paragraph just below eq.~\eqref{SWgen}, it is easy
to see that $A_\mu[a_\rho]$ is independent of $\phi$, while $\Phi$
is linear in $\phi$. This allows us to separate ${\cal  H}$ in
terms independent of $\phi$ and terms that are quadratic in
$\phi$, whose scaling behaviour is readily obtained just by using
dimensional analysis. Thus, we write ${\cal H}^{\text{st}}$ --"st"
stands for standard Seiberg-Witten map-- as an expansion in powers
of $h\theta^{\m\n}$ as follows
\begin{equation}
\begin{array}{l}
{\cal H}^{\text{st}}={\cal H}^{\text{st}}_A+{\cal
H}^{\text{st}}_\Phi,\,\, {\cal
H}^{\text{st}}_A=\Tr\int\!d^3\,\vec{x}\,B_iB_i,\,\,
{\cal H}^{\text{st}}_\Phi=\Tr\int\!d^3\,\vec{x}\,D_i\Phi D_i\Phi\\
{\cal H}^{\text{st}}_A=\sum_{l\geq0}h^l \,{\cal
H}^{\text{st}(l)}_A,\,\, {\cal
H}^{\text{st}}_\Phi=\sum_{l\geq0}h^l\, {\cal
H}^{\text{st}(l)}_\Phi.
\end{array}
\label{HAPhi}
\end{equation}
${\cal H}^{\text{st}}_{A}$ is independent of $\phi$, and
 ${\cal H}^{\text{st}}_{\Phi}$ is quadratic in $\phi$. The scaling
 properties of these terms are then given by
\begin{equation}
\delta {\cal H}^{\text{st}(l)}_{A}=(1+2l)\delta
\lambda{\cal H}^{\text{st}(l)}_{A},\,\,\delta {\cal H}^{\text{st}(l)}_{\Phi}=
(-1+2l)\delta\lambda{\cal H}^{\text{st}(l)}_{\Phi}.
\label{scaling}
\end{equation}
    When evaluating these terms in field configurations that can be
    written as in eq.~(\ref{expfields}), the following additional expansions are obtained:
\begin{equation*}
\begin{array}{l}
{\cal H}^{\text{st}(l)}_{A/\Phi}=\sum_{m\geq0}h^m \,
{\cal H}^{\text{st}(l,m)}_{A/\Phi},\quad{\cal H}^{\text{st}(l,m)}_{A/\Phi}=
\frac{1}{m!}\frac{d^m}{dh^m}{\cal H}^{\text{st}(l)}_{A/\Phi}[a_\mu^{(0)}
+h^k a_\mu^{(k) },\phi^{(0)}+h^k\phi^{(k)}]|_{h=0}.
\end{array}
\end{equation*}
    Therefore the invariance of ${\cal H}^{\text{st}}$
    under the infinitesimal transformations in eq.~(\ref{dilat})
is equivalent to:
\begin{equation*}
\sum_n h^n[(1+2n){\cal H}^{\text{st}(n)}_A+(-1+2n)
{\cal H}^{\text{st}(n)}_\Phi]=\!0\!=
\sum_{k\geq0} h^k\sum_{l=0}^k[(1+2l){\cal H}^{\text{st}(l,k-l)}_A+(-1+2l){\cal H}^{\text{st}(l,k-l)}_\Phi],
\end{equation*}
i.e.,
\begin{equation}
\sum_{l=0}^k[(1+2l){\cal H}^{\text{st}(l,k-l)}_A+(-1+2l){\cal H}^{\text{st}(l,k-l)}_\Phi]=0\,\,\forall k\geq0.
\label{derrick}
\end{equation}
    For $k=0$ this is equivalent to ${\cal H}^{\text{st}(0,0)}_A-{\cal H}^{\text{st}(0,0)}_\Phi=0$, which is satisfied
by the ordinary BPS monopoles. For $k=1$ eq.~(\ref{derrick}) gives
\begin{equation}
{\cal H}^{\text{st}(0,1)}_A-{\cal H}^{\text{st}(0,1)}_\Phi+3{\cal H}^{\text{st}(1,0)}_A+{\cal H}^{\text{st}(1,0)}_\Phi=0.
\label{derrick1}
\end{equation}
This relation holds trivially in the $SU(2)$ and $SO(5)$ cases
because, recalling that we are dealing with the standard
Seiberg-Witten applications, the first-order-in-$h\theta^{\m\n}$
contributions to the field configurations are just appropriate
linear combinations of the zero modes of the ordinary fields. It
is also satisfied in the $SU(3)$ case when evaluating in the field
configuration $\phi=\phi_\beta^{(0)}+h\phi^s T^s_\beta,\,
a_i=a_{i\beta}^{(0)}+ h a_i^s T^s_\beta$, with
$\phi_\beta^{(0)},\, a_{i\beta}^{(0)}$ given by
eqs.~(\ref{order0}) and (\ref{BPS}) and $\phi^s,\,a^s_i$ given in
eqs.~(\ref{sol1}) and (\ref{sol2}); each term in
eq.~(\ref{derrick1}) turns out to vanish, because all the traces are
of the type $\Tr\,T_\beta^s T_\beta^a=0$.\par
    By substituting $k=2$ in eq.~(\ref{derrick}), one obtains the following:
\begin{equation*}
 {\cal H}^{\text{st}(0,2)}_A-{\cal H}^{\text{st}(0,2)}_\Phi+
 3{\cal H}^{\text{st}(1,1)}_A+{\cal H}^{\text{st}(1,1)}_\Phi+5{\cal H}^{\text{st}(2,0)}_A+3{\cal H}^{\text{st}(2,0)}_\Phi=0.
\end{equation*}
    In contrast with the case analysed in ref.~\cite{Martin:2005vr},
    the equation involves the order $h^2\theta^2$ contributions
to the field configurations, due to the fact that in
eq.~\eqref{scaling} the ${\cal H}^{(n)}$ terms scale with powers
of $\lambda$ that are non-zero for $n=0$. Hence, we do not find
any obstruction --implied by Derrick's theorem-- to the existence,
at second-order in $h\theta^{\m\n}$ and for $a_0=0$, of static solutions to
the noncommutative Yang-Mills-Higgs equations.\par

    In the case of arbitrary SW maps, the scaling behaviour of the different
    contributions to the Hamiltonian is more complicated, due to the
fact that --see eq.~\eqref{SWgen}-- $A_i$ will receive
contributions with arbitrary even numbers of $\phi$'s. Therefore,
though we can always separate the terms of ${\cal H}$ as in
eq.~\eqref{HAPhi}, with  ${\cal H}_{\Phi}$ still scaling as in
eq.~\eqref{scaling}, now  ${\cal H}_{A}$ will not be independent
of $\phi$ and the scaling behaviour will change. Nevertheless, the
important issue is that there will exist terms ${\cal H}^{(n)}$
that will scale with powers of $\lambda$ that are non-zero for
$n=0$, so that when imposing the stationarity condition at order
$h^2\theta^2$, we will have again contributions of the type ${\cal
H}^{(0,2)}$ and the same conclusion as with the standard
Seiberg-Witten map will be reached.

\end{document}